\documentclass[longauth]{aa}

\usepackage{graphicx}
\usepackage{amsmath}	
\usepackage{amssymb}	
\usepackage{multicol}        
\usepackage{bm}		
\usepackage{pdflscape}	
\usepackage{threeparttable} 

\usepackage{rotating} 

\usepackage{xcolor}
\definecolor{darkgreen}{RGB}{0,100,0}
\usepackage[normalem]{ulem}
\usepackage{siunitx}

\usepackage{txfonts}

\def\sini{$\sin\,i$}              
\def\vsini{$v\,\sin\,i$}              
\def\kms{\hbox{\,km\,s$^{-1}$}}       
\def\vsini{\hbox{$v$\,sin\,$i$}}      
\def\sini{\hbox{sin\,$i$}}            

\newcommand{\teff}{$T_{\rm eff}$}
\newcommand{\logg}{$\rm log\,g_\star$}

\newcommand{\Tzerob}[1][days]{$7743.37545 _{ - 0.00051 } ^ { + 0.00051 } $#1} 
\newcommand{\Pb}[1][days]{$3.471745 _{ - 0.000046 } ^ { + 0.000044 } $#1}

\newcommand{\arb}[1][ ]{$10.43 _{ - 0.14 } ^ { + 0.14 } $#1} 
\newcommand{\rrb}[1][ ]{$0.03013 _{ - 0.00022 } ^ { + 0.00022 } $#1} 
\newcommand{\kb}[1][${\rm m\,s^{-1}}$]{$4.62 _{ - 0.65 } ^ { + 0.58 } $#1} 
\newcommand{\mpb}[1][$M_{\oplus}$]{$9.68 _{ - 1.37 } ^ { + 1.21 } $#1} 
\newcommand{\rpb}[1][$R_{\oplus}$]{$2.59 _{ - 0.06 } ^ { + 0.06 } $#1} 
\newcommand{\ib}[1][deg]{$86.846 _{ - 0.041 } ^ { + 0.041 } $#1} 
\newcommand{\ab}[1][AU]{$0.03817 _{ - 0.00092 } ^ { + 0.00095 } $#1} 
\newcommand{\insolationb}[1][${\rm F_{\oplus}}$]{$234.31 _{ - 18.28 } ^ { + 19.58 } $#1} 
\newcommand{\denstrb}[1][${\rm g\,cm^{-3}}$]{$1.782 _{ - 0.069 } ^ { + 0.071 } $#1} 
 
\newcommand{\denpb}[1][${\rm g\,cm^{-3}}$]{$3.07 _{ - 0.45 } ^ { + 0.45 } $#1} 
\newcommand{\grapb}[1][${\rm cm\,s^{-2}}$]{$1165.9 _{ - 169.9 } ^ { + 141.6 } $#1} 
\newcommand{\grapparsb}[1][${\rm cm\,s^{-2}}$]{$1420.1 _{ - 202.7 } ^ { + 191.8 } $#1} 
 
\newcommand{\Teqb}[1][K]{$1088.9 _{ - 21.9 } ^ { + 22.1 } $#1} 
\newcommand{\ttotb}[1][hours]{$2.180 _{ - 0.028 } ^ { + 0.029 } $#1}

\newcommand{\Tzeroc}[1][days]{$7744.53906 _{ - 0.00037 } ^ { + 0.00039 } $#1} 
\newcommand{\Pc}[1][days]{$7.138048 _{ - 0.000063 } ^ { + 0.000072 } $#1}

\newcommand{\arc}[1][ ]{$22.51_{ - 0.20 } ^ { + 0.20 } $#1} 
\newcommand{\rrc}[1][ ]{$0.0411476 _{ - 0.0003202 } ^ { + 0.0003004 } $#1} 
\newcommand{\kc}[1][${\rm m\,s^{-1}}$]{$5.90 _{ - 0.80 } ^ { + 0.86 } $#1} 
\newcommand{\mpc}[1][$M_{\oplus}$]{$15.68 _{ - 2.13 } ^ { + 2.28 } $#1} 
\newcommand{\rpc}[1][$R_{\oplus}$]{$3.53 _{ - 0.08 } ^ { + 0.08 } $#1} 
\newcommand{\ic}[1][deg]{$89.8610 _{ - 0.0012 } ^ { + 0.0012 } $#1} 
\newcommand{\ac}[1][AU]{$0.0824 _{ - 0.0018 } ^ { + 0.0018 } $#1} 
\newcommand{\insolationc}[1][${\rm F_{\oplus}}$]{$50.35 _{ - 3.83 } ^ { + 4.05 } $#1}

\newcommand{\denpc}[1][${\rm g\,cm^{-3}}$]{$1.95 _{ - 0.28 } ^ { + 0.32 } $#1} 
\newcommand{\grapc}[1][${\rm cm\,s^{-2}}$]{$1799.3_{ - 242.3 } ^ { + 260.7 } $#1} 
\newcommand{\grapparsc}[1][${\rm cm\,s^{-2}}$]{$1234.1 _{ - 172.4 } ^ { + 191.7 } $#1} 
 
\newcommand{\Teqc}[1][K]{$741.4 _{ - 14.5 } ^ { + 14.5 } $#1} 
\newcommand{\ttotc}[1][hours]{$2.520 _{ - 0.021 } ^ { + 0.022 } $#1}

\newcommand{\Tzerod}[1][days]{$7745.20100 _{ - 0.00073 } ^ { + 0.00076 } $#1} 
\newcommand{\Pd}[1][days]{$10.45582 _{ - 0.00023 } ^ { + 0.00025 } $#1}

\newcommand{\ard}[1][ ]{$32.20 _{ - 0.45 } ^ { + 0.45 } $#1} 
\newcommand{\rrd}[1][ ]{$0.0288961 _{ - 0.0002872 } ^ { + 0.0003220 } $#1} 
\newcommand{\kd}[1][${\rm m\,s^{-1}}$]{$0.68 _{ - 0.50 } ^ { + 0.77 } $#1} 
 
\newcommand{\rpd}[1][$R_{\oplus}$]{$2.48 _{ - 0.06 } ^ { + 0.06 } $#1} 
\newcommand{\id}[1][deg]{$89.6431 _{ - 0.0051 } ^ { + 0.0049 } $#1} 
\newcommand{\ad}[1][AU]{$0.1178 _{ - 0.0029 } ^ { + 0.0029 } $#1} 
\newcommand{\insolationd}[1][${\rm F_{\oplus}}$]{$24.61 _{ - 1.93 } ^ { + 2.07 } $#1}

\newcommand{\Teqd}[1][K]{$619.9 _{ - 12.5 } ^ { + 12.7 } $#1} 
\newcommand{\ttotd}[1][hours]{$2.504 _{ - 0.033 } ^ { + 0.035 } $#1}

\newcommand{\Tzeroe}[1][days]{$7741.8969 _{ - 0.0024 } ^ { + 0.0020 } $#1} 
\newcommand{\Pe}[1][days]{$14.76289 _{ - 0.00061 } ^ { + 0.00065 } $#1}

\newcommand{\are}[1][ ]{$49.43 _{ - 0.73 } ^ { + 0.41 } $#1} 
\newcommand{\rre}[1][ ]{$0.0227044 _{ - 0.0003501 } ^ { + 0.0003527 } $#1} 
\newcommand{\ke}[1][${\rm m\,s^{-1}}$]{$1.69 _{ - 0.72 } ^ { + 0.74 } $#1} 
 
\newcommand{\rpe}[1][$R_{\oplus}$]{$1.95 _{ - 0.05 } ^ { + 0.05 } $#1} 
\newcommand{\ie}[1][deg]{$89.7994 _{ - 0.0030 } ^ { + 0.0016 } $#1} 
\newcommand{\aeee}[1][AU]{$0.18041 _{ - 0.0043 } ^ { + 0.0042 } $#1} 
\newcommand{\insolatione}[1][${\rm F_{\oplus}}$]{$10.50_{ - 0.81 } ^ { + 0.87 } $#1}

\newcommand{\Teqe}[1][K]{$500.9 _{ - 9.9 } ^ { + 10.1 } $#1} 
\newcommand{\ttote}[1][hours]{$2.300 _{ - 0.019 } ^ { + 0.034 } $#1}

\newcommand{\Tzerof}[1][days]{$7985.84 _{ - 0.30 } ^ { + 0.30 } $#1} 
\newcommand{\Pf}[1][days]{$12.102 _{ - 0.056 } ^ { + 0.067 } $#1}

\newcommand{\kf}[1][${\rm m\,s^{-1}}$]{$12.29 _{ - 0.81 } ^ { + 0.80} $#1} 
 
\newcommand{\qone}[1][]{$0.272 _{ - 0.098 } ^ { + 0.092} $#1} 
\newcommand{\qtwo}[1][]{$0.62 _{ - 0.16 } ^ { + 0.22} $#1} 
\newcommand{\uone}[1][]{$0.646 _{ - 0.097 } ^ { + 0.081 } $#1} 
\newcommand{\utwo}[1][]{$-0.129 _{ - 0.158} ^ { + 0.170 } $#1} 
\newcommand{\CARMENES}[1][${\rm m\,s^{-1}}$]{$0.00130 _{ - 0.00068 } ^ { + 0.00071 } $#1} 
\newcommand{\HARPSN}[1][${\rm m\,s^{-1}}$]{$0.00168 _{ - 0.00076 } ^ { + 0.00081 } $#1}

\usepackage[colorlinks=true,citecolor=blue]{hyperref}
\usepackage{natbib}

\begin{document}

   \title{Detection and Doppler monitoring of EPIC 246471491, a system of four transiting planets smaller than Neptune}


   \author{E. Palle\inst{1,2}
   \and G. Nowak\inst{1,2}
   \and R.~Luque\inst{1,2}%
   \and D.~Hidalgo\inst{1,2}%
   \and O.~Barrag\'{a}n\inst{3}%
   \and J. Prieto-Arranz\inst{1,2}
   \and T.~Hirano\inst{4}%
   \and M.~Fridlund\inst{5,6}%
   \and D.~Gandolfi\inst{3}%
   \and J.~Livingston\inst{7}%
   \and F.~Dai\inst{8,9}%
   \and J.\,C.~Morales\inst{10,11}%
   \and M.~Lafarga\inst{10,11}%
  \and S.~Albrecht\inst{12}%
  \and R.~Alonso\inst{1,2}%
 \and P.~J.~Amado\inst{13}
 \and J.~A.~Caballero\inst{14}
 \and J.~Cabrera\inst{15}%
  \and W.\,D.~Cochran\inst{16}%
  \and Sz.~Csizmadia\inst{15}%
  \and H.~Deeg\inst{1,2}%
  \and Ph.~Eigm\"uller\inst{15,17}%
  \and M.~Endl\inst{18}%
  \and A.~Erikson\inst{15}%
  \and A.~Fukui\inst{19,1}%
\and E.W.~Guenther\inst{20}%
  \and S.~Grziwa\inst{21}%
  \and A.\,P.~Hatzes \inst{20}%
  \and J.~Korth\inst{21}%
  \and M.~K\"urster\inst{22}
  \and M.~Kuzuhara\inst{23,24}%
  \and P.~Monta\~nes Rodr\'iguez\inst{1,2}%
  \and F.~Murgas\inst{1,2}%
  \and N.~Narita\inst{23,24,1}%
  \and D.~Nespral \inst{1,2}%
  \and M.\,P\"atzold\inst{21}%
   \and C.\,M.~Persson \inst{6}%
  \and A.~Quirrenbach\inst{25}%
   \and H.~Rauer\inst{15,16,17}%
   \and S.~Redfield\inst{27}%
   \and A.~Reiners\inst{27}
   \and I.~Ribas\inst{10,11}%
  \and A.\,M.\,S.~Smith\inst{15}%
  \and V.~Van~Eylen\inst{5}%
  \and J.\,N.~Winn\inst{8}%
  \and M.~Zechmeister\inst{28}
}

\institute{
Instituto de Astrof\'\i sica de Canarias (IAC), 38205 La Laguna, Tenerife, Spain 
\email{jparranz@iac.es}
\and 
Departamento de Astrof\'\i sica, Universidad de La Laguna (ULL), 38206, La Laguna, Tenerife, Spain 
\and
Dipartimento di Fisica, Universit\`a di Torino, Via P. Giuria 1, I-10125, Torino, Italy 
\and 
Department of Earth and Planetary Sciences, Tokyo Institute of Technology, 2-12-1 Ookayama, Meguro-ku, Tokyo 152-8551, Japan 
\and
Leiden Observatory, Leiden University, 2333CA Leiden, The Netherlands 
\and
Department of Space, Earth and Environment, Chalmers University of Technology, Onsala Space Observatory, 439 92 Onsala, Sweden 
\and
Department of Astronomy, The University of Tokyo, 7-3-1 Hongo, Bunkyo-ku, Tokyo 113-0033, Japan 
\and 
Department of Astrophysical Sciences, Princeton University, 4 Ivy Lane, Princeton, NJ 08544, USA 
\and 
Department of Physics and Kavli Institute for Astrophysics and Space Research, Massachusetts Institute of Technology, Cambridge, MA 02139, USA 
\and 
Institut de Ci\`encies de l\'~Espai (ICE, CSIC), C/Can Magrans, s/n, Campus UAB, 08193 Bellaterra, Spain 
\and
Institut d’Estudis Espacials de Catalunya (IEEC), E-08034 Barcelona, Spain 
\and
Stellar Astrophysics Centre, Department of Physics and Astronomy, Aarhus University, Ny Munkegade 120, DK-8000 Aarhus C, Denmark 
\and
Instituto de Astrofísica de Andalucía (IAA-CSIC), Glorieta de la Astronomía s/n, E-18008 Granada, Spain
\and 
Centro de Astrobiología (CSIC-INTA), ESAC campus, Camino Bajo del Castillo s/n, E-28692 Villanueva de la Cañada, Madrid, Spain
\and
Institute of Planetary Research, German Aerospace Center, Rutherfordstrasse 2, 12489 Berlin, Germany 
\and
Institute of Geological Sciences, Freie Universit\"at Berlin, Malteserstr. 74-100, 12249 Berlin, Germany
\and 
Center for Astronomy and Astrophysics, TU Berlin, Hardenbergstr. 36, 10623 Berlin, Germany
\and 
Department of Astronomy and McDonald Observatory, University of Texas at Austin, 2515 Speedway, Stop C1400, Austin, TX 78712, USA 
\and 
Okayama Astrophysical Observatory, National Astronomical Observatory of Japan, NINS, Asakuchi, Okayama 719-0232, Japan 
\and 
Th\"uringer Landessternwarte Tautenburg, Sternwarte 5, 07778 Tautenburg, Germany 
\and 
Rheinisches Institut f\"ur Umweltforschung an der Universit\"at zu K\"oln, Aachener Strasse 209, 50931 K\"oln, Germany 
\and
Max-Planck-Institut für Astronomie, Königstuhl 17, D-69117 Heidelberg, Germany
\and     
Astrobiology Center, NINS, 2-21-1 Osawa, Mitaka, Tokyo 181-8588, Japan 
\and 
National Astronomical Observatory of Japan, NINS, 2-21-1 Osawa, Mitaka, Tokyo 181-8588, Japan 
\and
Landessternwarte, Zentrum für Astronomie der Universtät Heidelberg, Königstuhl 12, D-69117 Heidelberg, Germany
\and
Max-Planck-Institut f{\"u}r Astronomie, K{\"o}nigstuhl 17, 69117, Heidelberg, Germany 
\and
Astronomy Department and Van Vleck Observatory, Wesleyan University, Middletown, CT 06459, USA 
\and 
Institut für Astrophysik, Georg-August-Universität, Friedrich-Hund-Platz 1, D-37077 Göttingen, Germany
}

   \date{Received Month 00, 2017; accepted Month 00, 2017}

 
  \abstract
   {The Kepler extended mission, also known as \textit{{\it K2}}, has provided the community with a wealth of planetary candidates that orbit stars typically much brighter than the targets of the original mission. These planet candidates are suitable for further spectroscopic follow-up and precise mass determinations, leading ultimately to the construction of empirical mass-radius diagrams. Particularly interesting is to constrain the properties of planets between the Earth and Neptune in size, the most abundant type of planets orbiting Sun-like stars with periods less than a few years.}
   {Among many other \textit{{\it K2}} candidates, we discovered a multi-planetary system around EPIC~246471491, with four planets ranging in size from twice the size of Earth, to nearly the size of Neptune. We aim here at confirming their planetary nature and characterizing the properties of this system.}
  {We measure the mass of the planets of the EPIC~246471491 system by means of precise radial velocity measurements using the CARMENES spectrograph and the HARPS-N spectrograph.}
   {With our data we are able to determine the mass of the two inner planets of the system with a precision better than 15\%, and place upper limits on the masses of the two outer planets.}
   {We find that EPIC~246471491~b has a mass of $M_\mathrm{b}$\,=\,\mpb\, and a radius of $R_\mathrm{b}$\,=\,\rpb\,, yielding a mean density of $\rho_\mathrm{b}$\,=\,\denpb, while EPIC~246471491~c has a mass of $M_\mathrm{c}$\,=\,\mpc\,, radius of $R_\mathrm{c}$\,=\,\rpc\,, and a mean density of $\rho_\mathrm{c}$\,=\,\denpc. For EPIC~246471491~d ($R_\mathrm{d}$\,=\,\rpd\,) and EPIC~246471491~e ($R_\mathrm{e}$\,=\,\rpe\,) the upper limits for the masses are 6.5\,$M_\oplus$ and 10.7\,$M_\oplus$, respectively. The system is thus composed of a nearly Neptune-twin planet (in mass and radius), two sub-Neptunes with very different densities and presumably bulk composition, and a fourth planet in the outermost orbit that resides right
in the middle of the super-Earth/sub-Neptune radius gap. Future comparative planetology studies of this system can provide useful insights into planetary formation, and also a good test of atmospheric escape and evolution theories.}


\titlerunning{Four planets on EPIC~246471491}

   \maketitle
%
\section{Introduction}
\label{sec:intro}

Space-based transit surveys such as CoRoT \citep{2009A&A...506..411A} and Kepler \citep{2010Sci...327..977B} have revolutionized the field of exoplanetary science. Their high-precision and nearly uninterrupted photometry has opened the doors to explore planet parameter spaces that are not easily accessible from the ground, most notably, the Earth-radius planet domain. However, our knowledge of both super-Earths ($R_{p}$ = 1--2\,$R_\oplus$ and $M_{p}$ = 1--10\,$M_\oplus$) and Neptune planets ($R_{p}$ = 2--6\,$R_\oplus$ and $R_{p}$ = 10--40\,$M_\oplus$) is still limited, due to the small radial velocity (RV) variation induced by such planets and the relative faintness of most of {\it Kepler} host stars ($V>13$\,mag) which make precise mass determinations difficult.

Thus, many questions remain unanswered, for example what is the composition and internal structure of small planets?
\citet{Fulton17} and \citet{Fulton18} reported a radius gap at $\sim 2\,R_\oplus$ in the exoplanet radius distribution using {\it Kepler} data for Sun-like stars, and \citet{2018AJ....155..127H} indicated that the gap could extend down to the M dwarf domain. This would point to a very different planetary nature for planets on each side of the gap. Is this due to planet migration? Are the larger planets surrounded by a H/He atmospheres while the smaller planet have lost these envelopes? Or, did they already form with very different bulk densities? Answering these questions requires statistically significant samples of well-characterized small planets, especially in terms of orbital parameters, mass, radius and mean density.

{\it Kepler}'s extended {\it K2} mission is a unique opportunity to gain knowledge about small close-in planets. Every 3 months, {\it K2} observes a different stellar field located along the ecliptic, targeting up to 15 times brighter stars than the original {\it Kepler} mission. The KESPRINT collaboration\footnote{\href{http://www.iac.es/proyecto/kesprint}{http://www.iac.es/proyecto/kesprint}} is an international effort dedicated to the discovery, confirmation and characterization of planet candidates from the space transit missions {\it K2} and {\it TESS} and, in the future, {\it PLATO}. We have been focusing on determining the masses of small planets around bright stars, especially for planets in or around the radius gap.

Here, we present the discovery and characterization of four transiting planets around the star EPIC~246471491. While these planets are observed to have radii between 2 and 3.5 radii of the Earth, our follow-up observations indicate that they have very different bulk compositions. This has significant implications for the physical nature of planets around the radius gap. In this paper we provide ground-based follow-up observations that confirm that EPIC~246471491 is a single object and establish it main stellar properties. We also analyze jointly the {\it K2} data together with high-precision RV data from CARMENES and HARPS-N spectrographs, to retrieve orbital solutions and planetary masses. Finally we discuss the possible bulk compositions of the planets, leading to different densities.



\section{{\it K2} photometry and candidate detection}
\label{sec:photo}

EPIC~246471491 (RA = 23:17:32.23, DEC = 01:18:01.04, in the Aquarius constellation) was proposed as a  {\it K2} GO target for Campaign 12  in several programs (GO-12123, PI Stello; GO-12049, PI Quintana; and GO-12071, PI Charbonneau). The star was observed for 78.85 days from 15th December 2016 to 4th March 2017. During this interval, the {\it Kepler} spacecraft entered safe mode from 1st to 6th February 2017, causing a gap of 5.3 days in the data. 

\subsection{Light curve extraction and planet detection}

\begin{figure}
\centering
\includegraphics[width=8.3cm,angle=0,clip=true]{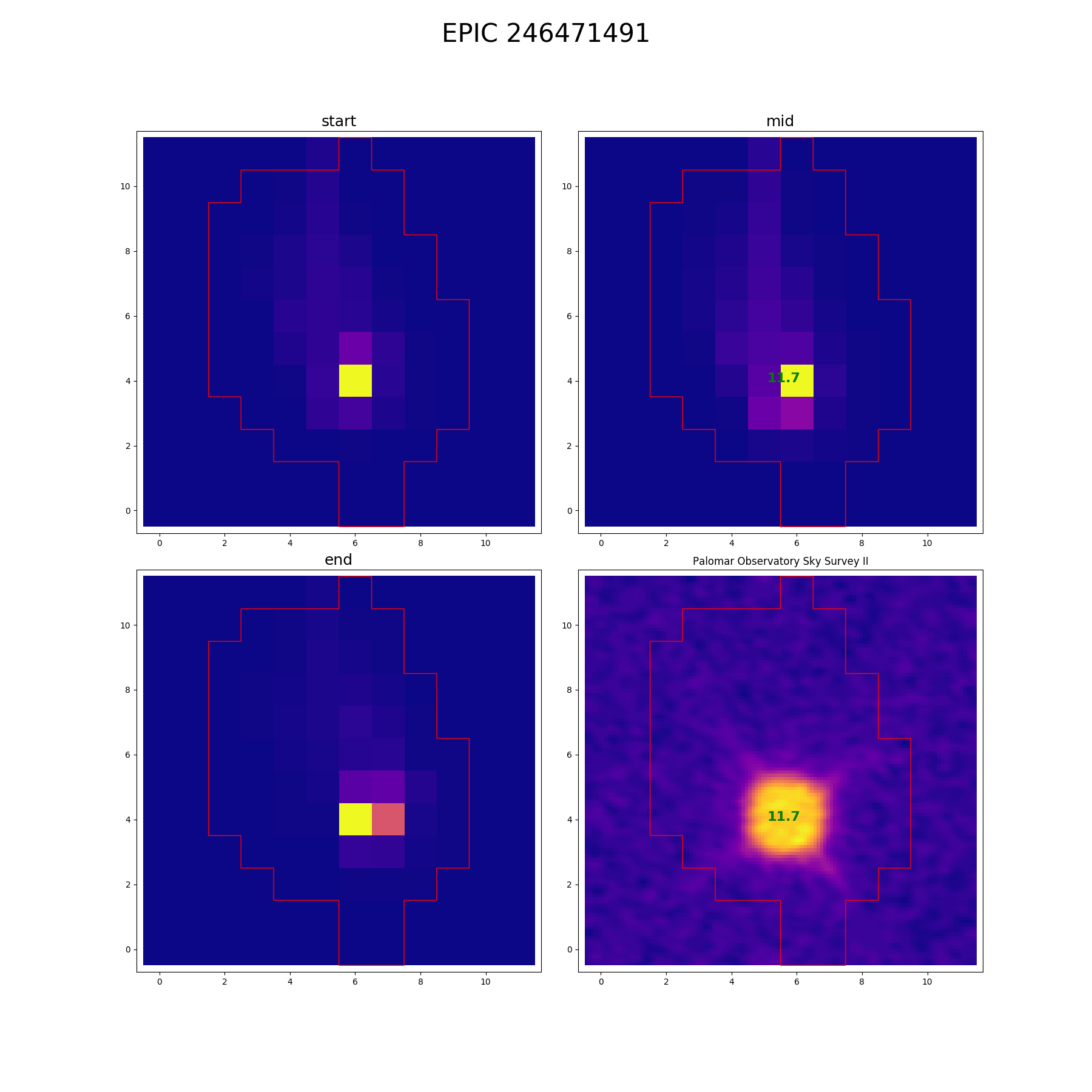}
\caption{{\it K2} image of the object EPIC~246471491. A custom-built changing aperture is fitted based on the pixel counts of the star and background. The image shows three typical apertures used at the beginning ({\it top left}), mid ({\it top right}) and end ({\it bottom left}) of the time series. The bottom right panel shows a high resolution image of the same field taken from Palomar Observatory.}
\label{aperture}
\end{figure}

We built the light curve of EPIC~246471491 directly from raw data (files downloaded from the Mikulski Archive for Space Telescopes\footnote{\href{https://archive.stsci.edu/kepler/data_search/search.php}{https://archive.stsci.edu/kepler/data\_search/search.php}}, MAST), using the long cadence (LC) version (29.4 min time stamps). Our pipeline is based on the implementation of the pixel level decorrelation (PLD) model \citep{2015ApJ...805..132D}, and a modified version of the Everest\footnote{\href{https://github.com/rodluger/everest}{https://github.com/rodluger/everest}} pipeline \citep{2017arXiv170205488L}. 
The PLD model uses a Taylor expansion of the instrumental signal as regressors in a linear model. These regressors are the products of the fractional fluxes in each pixel of the target aperture. The optimal aperture is built by searching for the photo-center and selecting pixels with a threshold of $1.2\sigma$ over the previously calculated background  (Figure~\ref{aperture}). The pipeline extracts the raw light curve from the apertures, removing time cadences with bad quality flags, and the background contribution. Next, it fits a regularized regression model to the data, iteratively up to the third order, and applies the cross-validation method to obtain the regularization matrix coefficients and Gaussian processes to compute the covariance matrix. All these steps are described in \citep{2017arXiv170205488L}. 

\begin{figure}
\centering
\includegraphics[width=8.3cm,angle=0,clip=true]{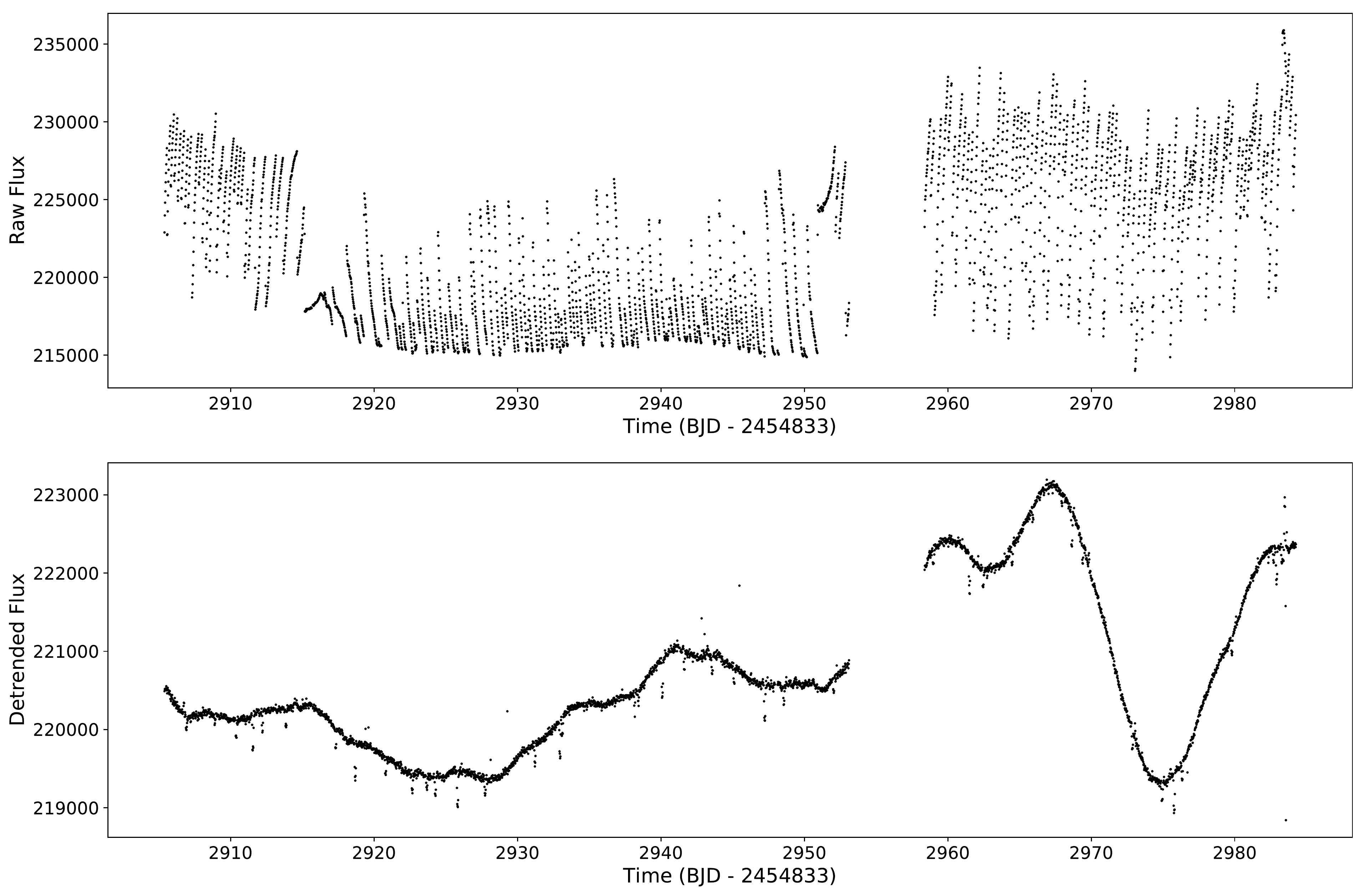}
\caption{Kepler light curves of EPIC~246471491. {\it Top}: Original raw light curves as derived from raw flux data. {\it Bottom}: De-trended light curve after analysis with our modified Everest-based pipeline. Stellar variability of the order of tens of days and the transits of several planets are clearly visible.}
\label{lightcurve}
\end{figure}

Prior to planet searches, we need to flatten the  \textit{{\it K2}} light curve by applying a robust locally weighted regression method \citep{Cleveland1979} iteratively until no outliers are detected. We remove 3$\sigma$ outliers replacing these points by the median of the neighbors. Note that the first two days and the last day of data, which shows anomalies probably related to thermal settling, were removed from our analysis. Applying these method iteratively we are able to remove any stellar flares. We then divide the original light curve by this variability model. The initial and final de-trended \textit{{\it K2}} light curves are plotted in Figure~\ref{lightcurve}.

Next we perform a Box-fitting Least Squares (BLS) algorithm \citep{2002A&A...391..369K} to detect the exact period of each possible planetary signal in the light curve. The BLS algorithm is very sensitive to outliers, therefore, we remove them by performing a sigma clipping. In this case, a value of  30\,$\sigma$ is enough. Once a planetary signal is detected in the power spectrum, we remove that specific transit signal by applying the BLS algorithm iteratively until no further signals are detected. 

\begin{figure}
\centering
\includegraphics[width=8.3cm,angle=0,clip=true]{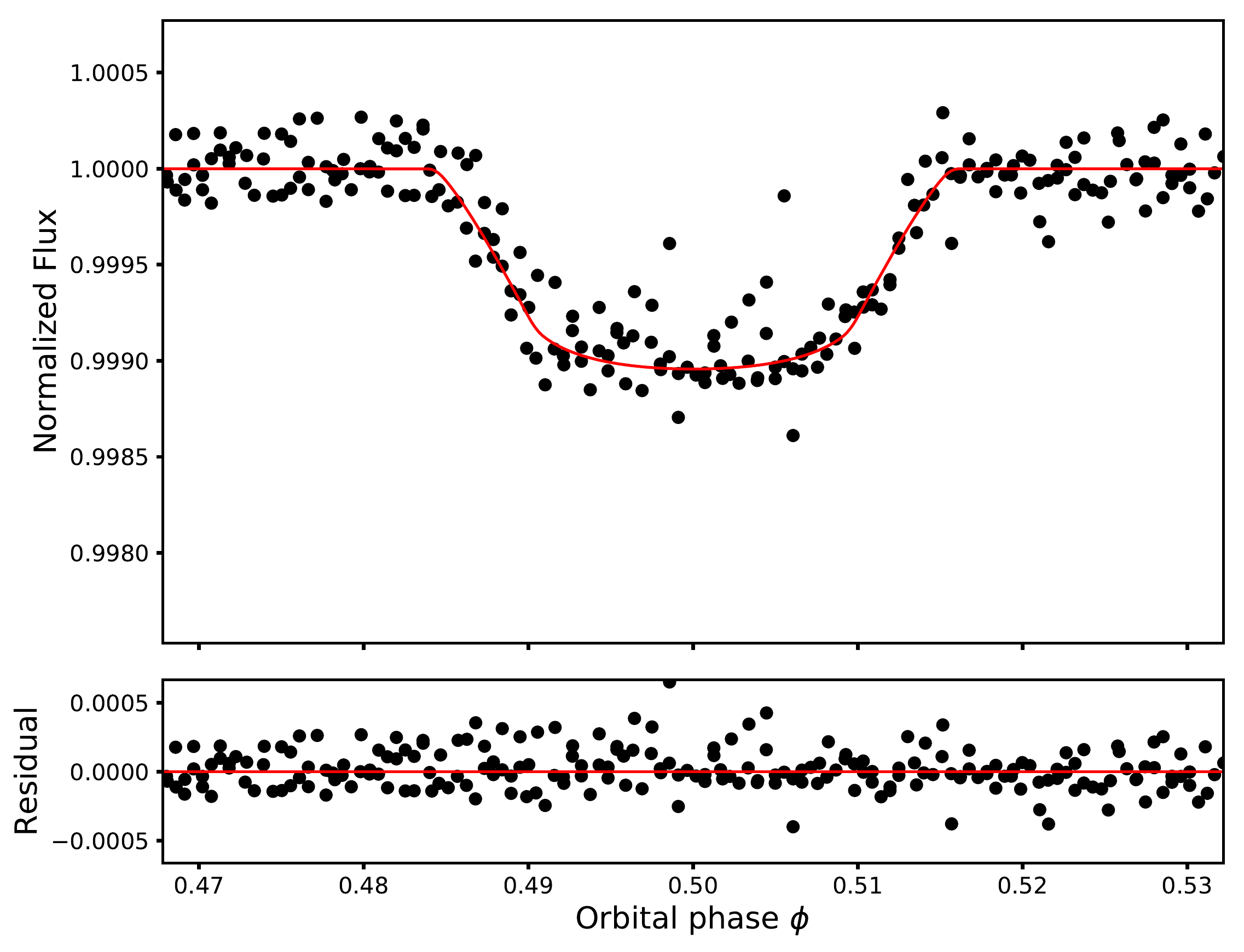}
\includegraphics[width=8.3cm,angle=0,clip=true]{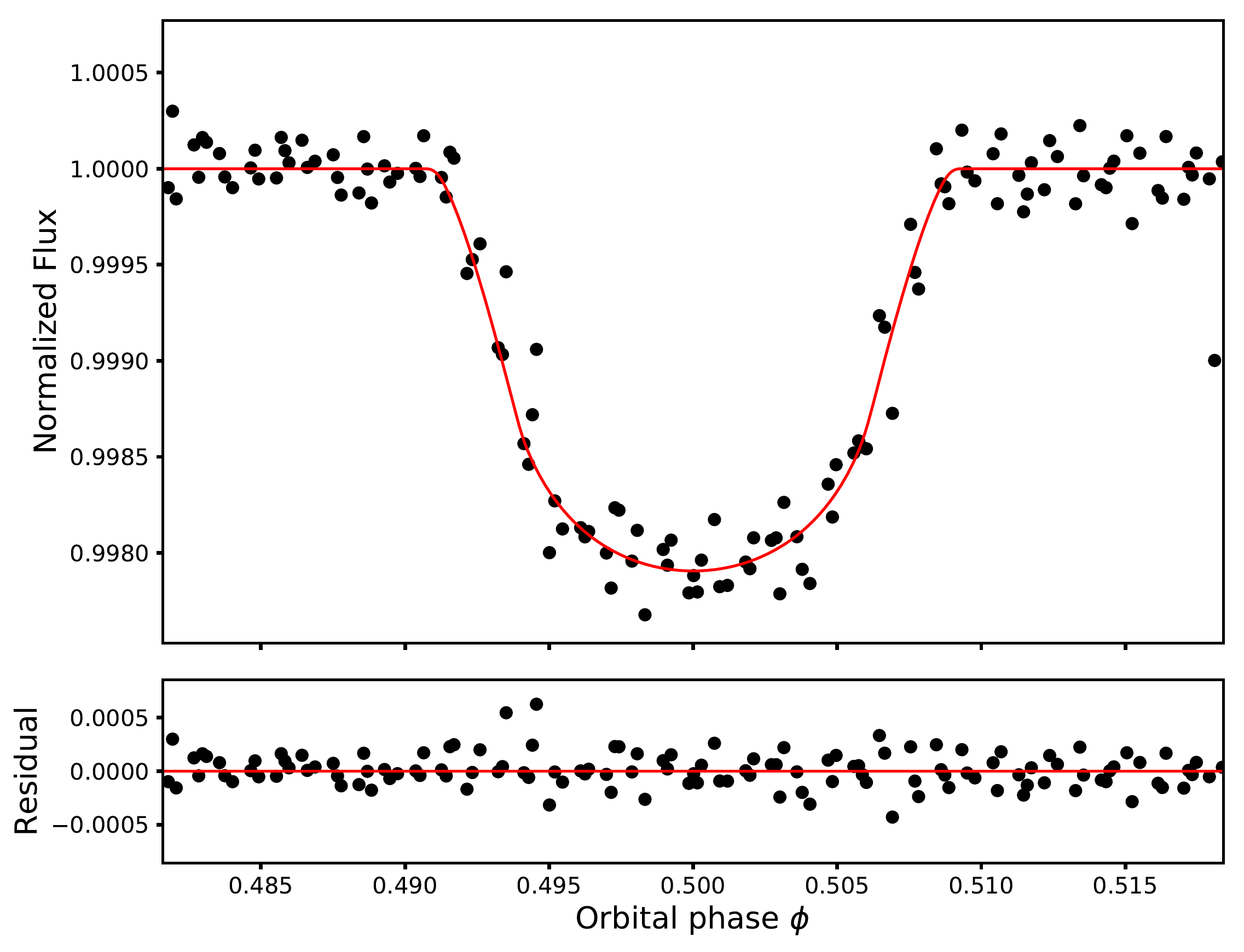}
\includegraphics[width=8.3cm,angle=0,clip=true]{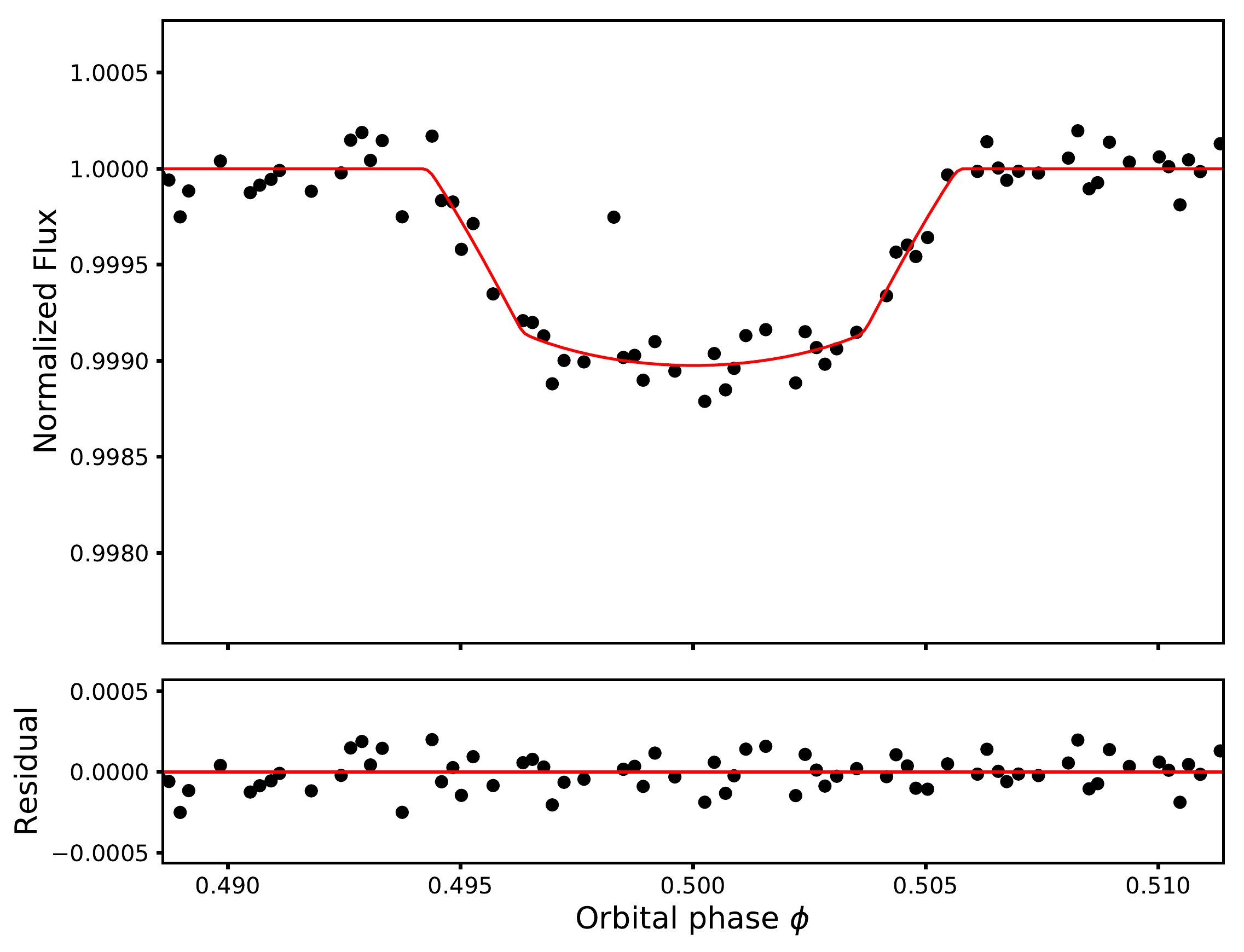}
\includegraphics[width=8.3cm,angle=0,clip=true]{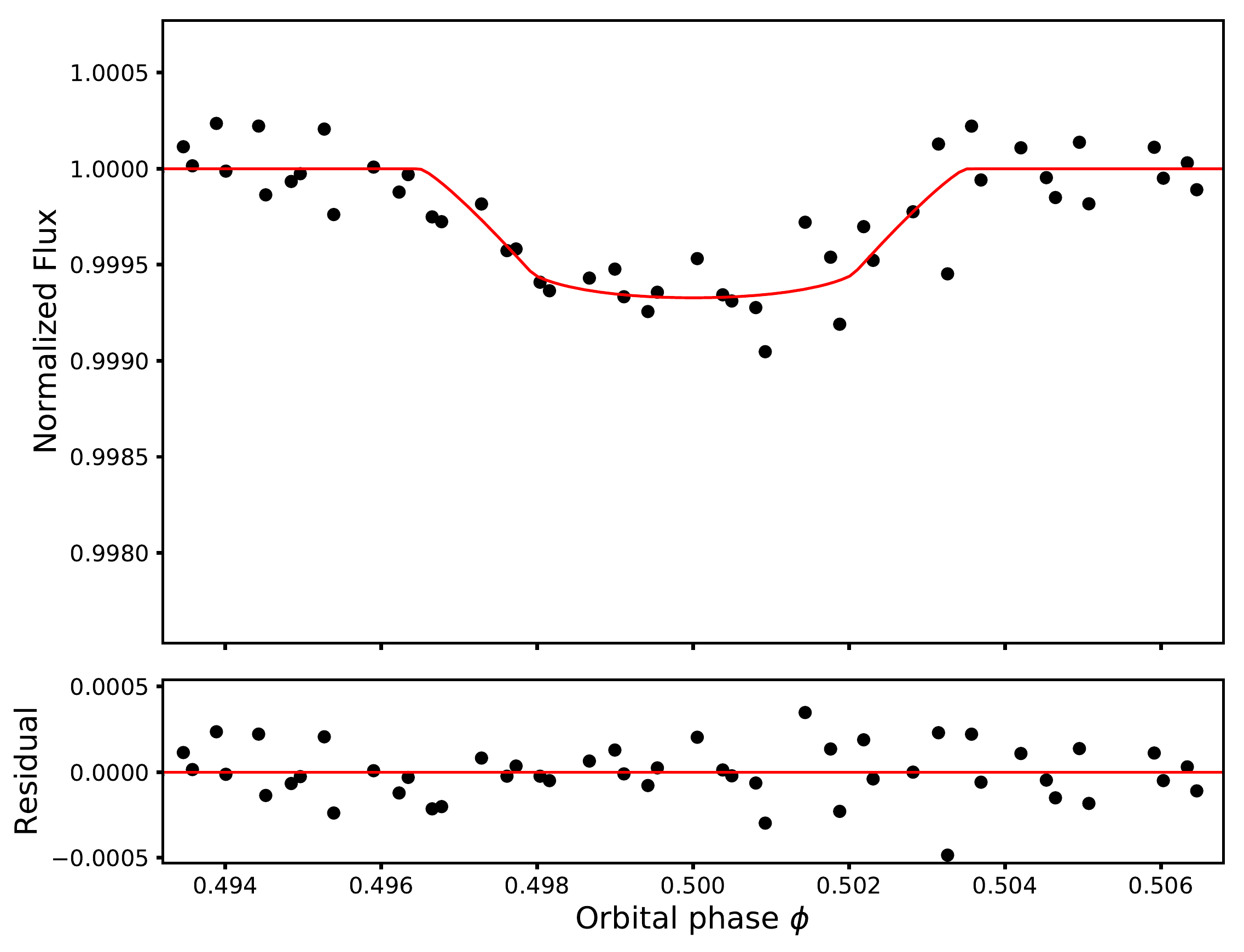}
\caption{Phase-folded transit light curves of the four planets detected orbiting EPIC~246471491. {\it Top panels}: transit light curves and best-fit transit models (red) on the same flux scale. {\it Lower panels}: residuals of the transit fit. }
\label{transits}
\end{figure}

Four planet detections were made in the course of the analysis of EPIC~246471491, at periods of 3.47, 7.13, 10.45 and 14.76 days (see Figure~\ref{transits}). The planet periods are close to 1\,:\,2\,:\,3\,:\,4 commensurability, but not quite, being the real ratio numbers 1\,:\,2.05\,:\,3.01\,:\,4.25. This near-commensurability may be indicative that the system is in resonance. Figure~\ref{transits} shows the phase-folded light curve for each transit, and its best-fit transit model. We fit every transit individually with the \texttt{python} package \texttt{batman} \citep{2015PASP..127.1161K}. We tentatively fit every transit with a non-linear least-squares minimization routine yielding good results for the transit parameters and taking these as input for a fine fitting with the MCMC method implemented in \texttt{emcee} \citep{2013PASP..125..306F}, using 100 walkers and 30000 steps. We then remove the first 22\,500 steps to estimate the uncertainties in the transit parameters. Once we obtain the MCMC results for the transit parameters of a planet, we iteratively remove the points inside the transit for the next fitting. The retrieved planetary parameters derived from fitting the \textit{{\it K2}} light curve alone are given in Table~\ref{planetparams}. 

The auto-correlation analysis of the \textit{{\it K2}} photometry retrieves a stellar rotational period at around 17 days, but the auto-correlation peak is broad and non-significant. We discuss this point further in Section~\ref{sec:results}.


\section{Ground-based follow-up observations}
\label{sec:obs}

\subsection{Lucky imaging and AO observations}

\begin{figure}
\includegraphics[width=8.7cm,angle=0,clip=true]{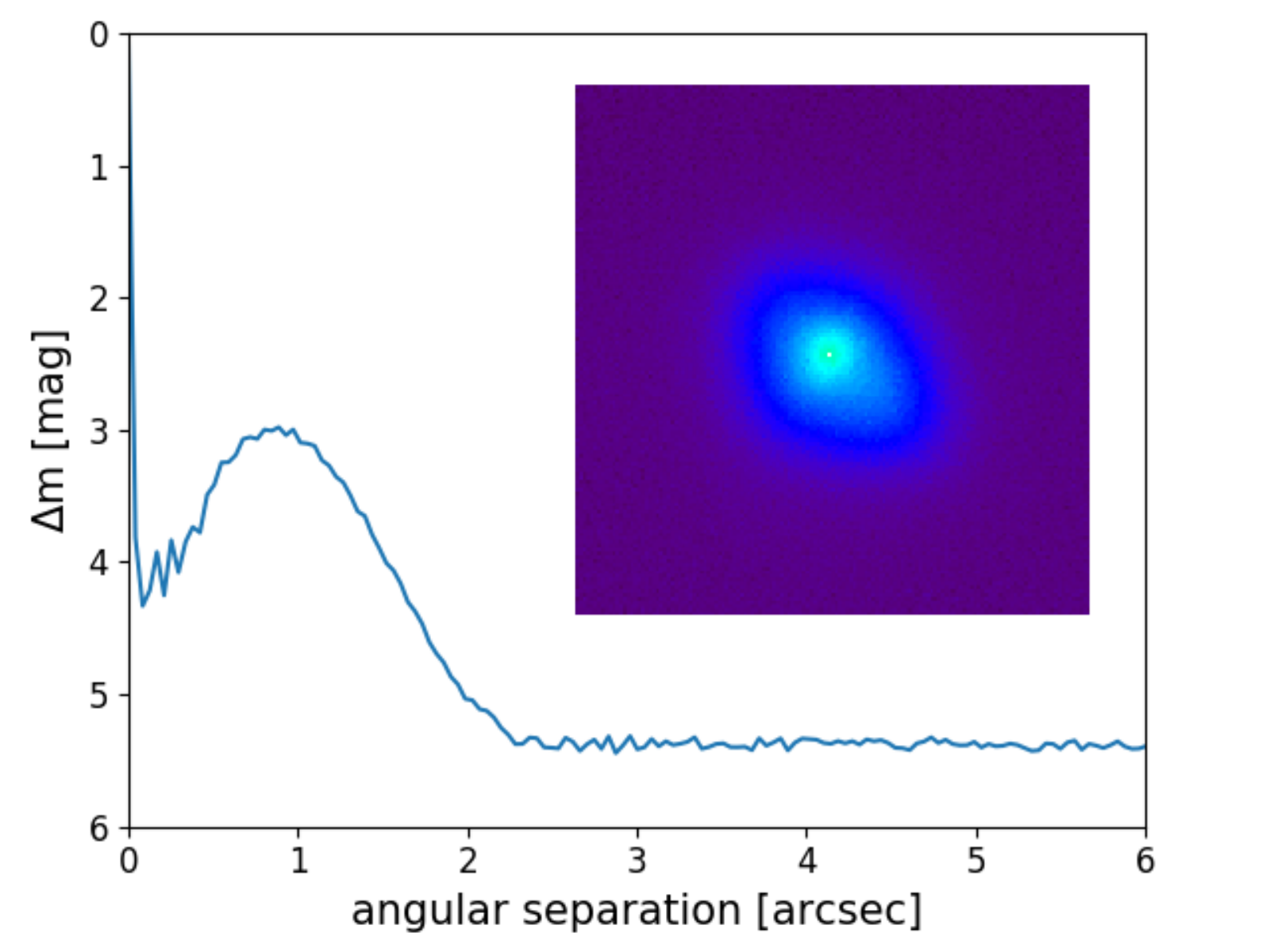}
\includegraphics[width=8.3cm,angle=0,clip=true]{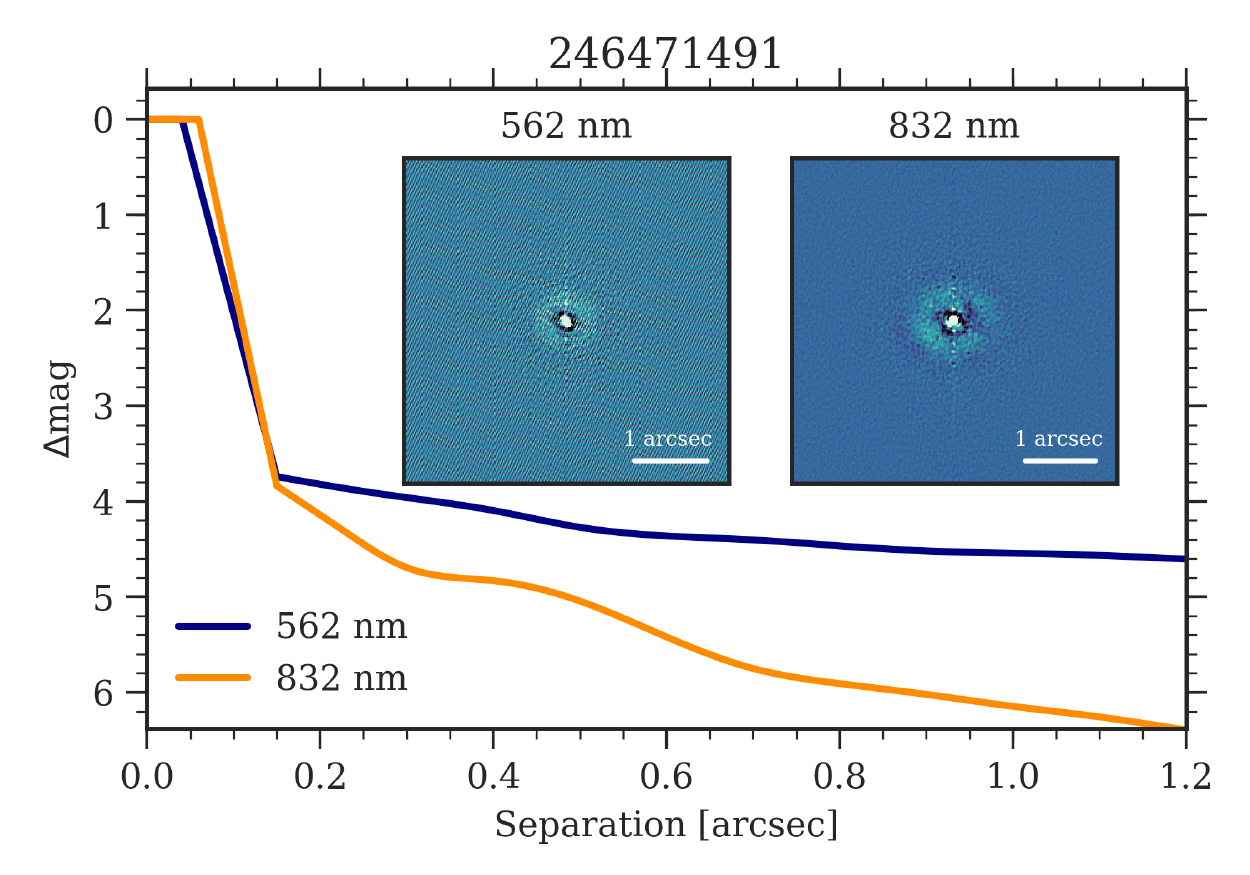}
\includegraphics[width=8.0cm,angle=0,clip=true]{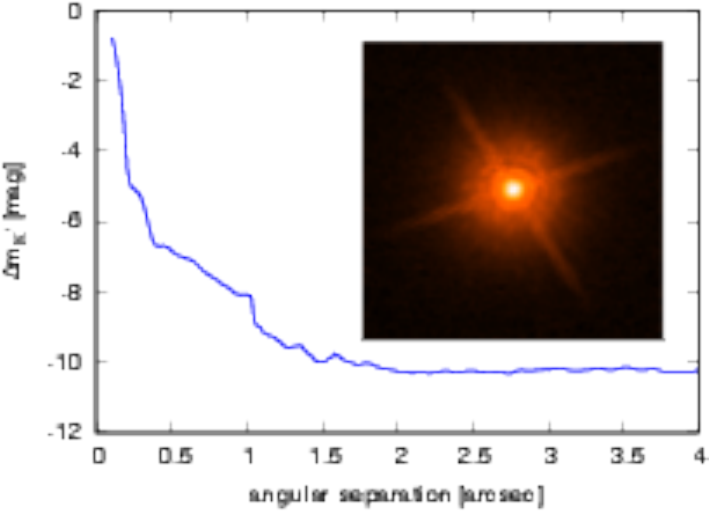}
\caption{{\it Top}: $I$-band magnitude contrast curve as a function of angular separation up to $6.0\arcsec$ from EPIC~246471491 obtained with the FastCam camera at TCS. The solid line indicates the $5\sigma$ detection limit. The inset shows the $6\arcsec \times 6\arcsec$ combined image of EPIC~246471491. North is right and East is down. {\it Middle}: Contrast curves at 562\,nm and 832\,nm obtained with NESSI. Inset images at both wavelengths are also shown, with a FOV of $4.6\arcsec \times 4.6\arcsec$. {\it Bottom}: Subaru AO image (inset) and $5\sigma$ contrast light curve of EPIC~246471491. The inset image has a FOV of $4\arcsec\times4\arcsec$.}
\label{lucky}
\end{figure}

We performed Lucky Imaging (LI) of EPIC~246471491 with the FastCam camera \citep{oscoz08} at the 1.52-m Telescopio Carlos S\'anchez (TCS). FastCam is a very low noise and fast readout EMCCD camera with $512 \times 512$ pixels (with a physical pixel size of 16 microns and a FoV of $21.2\arcsec \times 21.2\arcsec$). On the night of UT 2018 July 15, 10\,000 individual frames of EPIC~246471491 were collected in the Johnson-Cousins $I$-band, with an exposure time of 50\,ms for each frame. Figure \ref{lucky} shows the contrast curve that was computed based on the scatter within the annulus as a function of angular separation from the target centroid (see \citet{jorge} for details). We used a high resolution image constructed by co-addition of the $30\%$ best images, so that it had a 150\,s total exposure time. No bright companion was detectable within $6.0\arcsec$.

On the night of UT 2018 June 19, we also observed EPIC~246471491 with the NASA Exoplanet Star and Speckle Imager (NESSI, \citet{2016SPIE.9907E..2RS, 2018PASP..130e4502S}) on the 3.5-m WIYN telescope at the Kitt Peak National Observatory. NESSI uses electron-multiplying CCDs to conduct speckle-interferometric imaging, capturing a series of 40\,ms exposures simultaneously at 562\,nm and 832\,nm. The data were acquired and reduced following the procedures described by \citet{2011AJ....142...19H}, yielding reconstructed $4.6\arcsec \times 4.6\arcsec$ images. No secondary sources were detected and contrast curves were produced using a series of concentric annuli centered on the target. The reconstructed images achieve a contrast of $\sim 4$\,mag at $0\farcs2$ (see Figure~\ref{lucky}), which strongly constrains the possibility that the observed transit signals come from a nearby faint star. For more details on the use of NESSI for exoplanet validation and host star binarity determination, see \citet{2018arXiv180611504L} and \citet{2018AJ....156...31M}.

Finally, we also obtained high-resolution images for EPIC~246471491 using the InfraRed Camera and Spectrograph \citep[IRCS:][]{2000SPIE.4008.1056K}
and the adaptive-optics (AO) system on the Subaru 8.2-m telescope on UT 2018 June 14. To check for the absence of nearby companions, we imaged the target in the $K^\prime$ band with the fine-sampling mode ($1\,\mathrm{pix}=20\,\mathrm{mas}$), and implemented two types of sequences with a five-point dithering. For the first sequence we used a neutral-density (ND) filter whose transmittance is $\sim 0.81\%$ in the $K^\prime$ band to obtain unsaturated frames for the absolute flux calibration. We then acquired saturated frames to look for faint nearby companions. The total integration times amounted to 450\,s and 45\,s for the unsaturated and saturated frames, respectively. We reduced and median-combined those frames following the procedure described in \citet{2016ApJ...820...41H}. The combined images revealed no nearby companion around EPIC~246471491. To check for the detection limit, we drew the $5\sigma$ contrast curve following \citet{2016ApJ...820...41H} based on the combined saturated image. As plotted in Figure~\ref{lucky} , $\Delta m_{K^\prime}=7$ was achieved at $\sim 0\farcs5$ from the target. The inset of the figure displays the target image with field-of-view of $4\arcsec \times 4\arcsec$.

\subsection{CARMENES RV observations}

Radial velocity measurements of EPIC~246471491 were taken with the CARMENES spectrograph, mounted at the 3.5-m telescope at the Calar Alto Observatory in Spain. The CARMENES instrument has two arms \citep{CARMENES}, the visible (VIS) arm covering the spectral range $0.52$--$\SI{0.96}{\micro\metre}$ and a near infrared (NIR) arm covering the $0.96$--$\SI{1.71}{\micro\metre}$ range.  Here we use only the VIS channel observations to derive radial velocity measurements. The overall instrumental performance of CARMENES has been described by \citet{Reiners18}.

A total of 29 measurements were taken over the period 2017 September 20 to 2017 December 17, covering a time span of 98 days. In all cases exposure times were set at 1800\,s. Radial velocity values, chromatic index (CRX), differential line width (dLW) and H$\alpha$ index were obtained using the SERVAL program \citep{SERVAL}. For each spectrum, we also computed the cross-correlation function (CCF) and its full width half maximum (FWHM), contrast (CTR), and bisector velocity span (BVS) following \citet{Reiners18}. The RV measurements are given in Table~\ref{rvs_carmenes}, corrected for barycentric motion, secular acceleration and nightly zero-points. For more details see \citet{Trifonov18} and Luque et al (2018).

\subsection{HARPS-N RV observations}

Radial velocity measurements were also taken with the HARPS-North spectrograph, mounted at the 3.5-m TNG telescope at the Roque de los Muchachos Observatory in Spain. The HARPS-N instrument \citep{2012SPIE.8446E..1VC} covers the spectral range from $0.383$--$\SI{0.693}{\micro\metre}$. In total, 9 HARPS-N measurements were taken over the period 2017 September 16 to 2018 January 10, covering a time span of 112 days. Exposure times were set at 3600\,s. To derive radial velocities, SERVAL was also applied to the data. The performance of SERVAL RV extraction compared to the standard HARPS and HIRES pipelines is described in \citet{Trifonov18}. Both CARMENES and HARPS-N radial velocity measurements are given in Table~\ref{rvs_carmenes}.

\section{Host star properties}
\label{sec:star}

\begin{table}[!t]
  \centering 
  \caption{Stellar parameters of EPIC~246471491.}
\label{table_star}
  {\renewcommand{\arraystretch}{1.2}
  \begin{tabular}{lr}
    \hline
    \hline
    \multicolumn{2}{c}{EPIC~246471491} \\
    \hline
    RA$^1$ (J2000.0)   &   23:17:32.23 \\
	DEC$^1$ (J2000.0)  &  01:18:01.04 \\
	$V$-band magnitude$^2$ (mag)  &  $12.030 \pm 0.121$\\
	Spectral type$^2$  &   K2\,V\\
	Effective temperature$^2$ \teff~(K)  &  $4975\pm95$\\
    Surface gravity$^2$ \logg~(cgs)  &  $4.4\pm0.1$\\
    Iron abundance$^2$ [Fe/H]~(dex)  &  $0.0\pm0.05$\\  
    Mass$^2$ $M_\star$ ($M_{\odot}$)  &  $0.830\pm 0.023$\\
    Radius$^2$ $R_\star$ ($R_{\odot}$)  &  $0.787\pm 0.016$\\
    Projected rot. velocity$^2$ \vsini~(\kms)  &  $3.9\pm0.8$ \\
    Microturbulent velocity$^3$ $v_\mathrm{mic}$ (\kms) &  $0.82$ (fixed) \\
    Macroturbulent velocity$^4$ $v_\mathrm{mac}$ (\kms) &  $2.5$ (fixed)\\
    Interstellar reddening$^2$ $A_\mathrm{v}$ (mag) & $0.07\pm0.01$ \\
	Distance$^5$ (pc)  &  $155.6\pm6.4$\\
    \hline
  \end{tabular}}
\label{starpar}
\\
\flushleft
$^1$ {\it Hipparcos}, the New Reduction \citep{2007A&A...474..653V}. \\
$^2$ This work and AAVSO (https\://www.aavso.org/). \\
$^3$ \citet{2010MNRAS.405.1907B} \\
$^4$ \citet{2014MNRAS.444.3592D} \\
$^5$ {\it Gaia} DR2 \citet{2018arXiv180409366L} 
\end{table}

To retrieve the physical properties of EPIC~246471491, we analysed the co-added, radial velocity corrected, CARMENES spectra using the Spectroscopy Made Easy (SME) code \citep{2017A&A...597A..16P}. SME is designed to derive the fundamental parameters of stars. It iteratively calculates the synthesized spectrum based on a large grid of model spectra. The synthesized spectrum is fitted to the observed spectra using a $\chi^2$ minimization process. In this case, we used 1-D MARCS models \citep{2008A&A...486..951G}. Providing the code with fixed turbulent velocities $v_\mathrm{mac}=2.5\pm0.7$ \kms \citep{2014MNRAS.444.3592D} and $v_\mathrm{mic}=0.82\pm0.4$ \kms \citep{2010MNRAS.405.1907B}, we solved for $T_\mathrm{eff}$ by analyzing the Balmer profile of H$\alpha$, \logg~by fitting the Ca~I triplet at 6102, 6122 and 6162 \AA, and [Fe/H] and \vsini~by fitting $\sim 50$ Fe lines. We find $T_\mathrm{eff}=4975\pm95$\,K, \logg\,$=4.4\pm0.1$\,dex, $\mathrm{[Fe/H]}=0.00\pm 0.05$\,dex and \vsini\,$=3.9\pm0.8$ \kms, respectively. See Table~\ref{table_star} for a summary of EPIC~246471491 stellar parameters.

We confirmed the effective temperature and the value for \logg~  by also modeling the Na~I doublet (5889.95/5895.9\AA), using SME,  and deriving the abundance for Na~I from fainter lines in our CARMENES spectrum. Also, by analyzing the equivalent width of the interstellar sodium components \citep{2012MNRAS.426.1465P}, we find an extinction of $\mathrm{E}(B-V)=0.02\pm0.003$ that corresponds to $A_V = 0.07\pm0.01$\,mag.

We also used the HARPS-N co-added spectrum to derive stellar parameters. In particular, we fitted the spectral energy distribution using low-resolution model spectra with the same spectroscopic parameters as those found using the CARMENES co-added spectrum. Our results return an interstellar reddening value of $A_V = 0.1\pm0.05$\,mag.

We then used the $T_\mathrm{eff}$ and [Fe/H] values retrieved by SME, along with the new {\it Gaia} parallax value of $\pi =6.43 \pm0.11$\,mas \citet{2018arXiv180409366L}.  We quadratically added 0.1 mas to {\it Gaia}'s nominal uncertainty to account for systematics \citep[see][]{2018arXiv180409376L}. 

The stellar magnitude in $V$ band is taken from the AAVSO Photometric All Sky Survey (APASS) and corrected for extinction. 
The PARAM\footnote{\href{https://stev.oapd.inaf.it}{https://stev.oapd.inaf.it}} models \citep{2006A&A...458..609D} returns a stellar mass of  $M_\star = 0.830\pm0.023\,M_{\odot}$, a radius of $R_\star= 0.787\pm0.016\,R_{\odot}$ and a \logg\,$=4.539\pm0.024$\,cgs. The latter value of surface gravity is consistent with the SME value within less than $2\sigma$. As a sanity check, we used the bolometric correction from \citet{2010A&ARv..18...67T} and got a radius of $R_\star=0.880\pm0.080\,R_{\odot}$,  which is roughly consistent with the previous value. 

\section{Frequency analysis of RV and photometric data}
\label{sec:results}

\begin{figure}
\centering
\includegraphics[width=8.3cm,angle=0,clip=true]{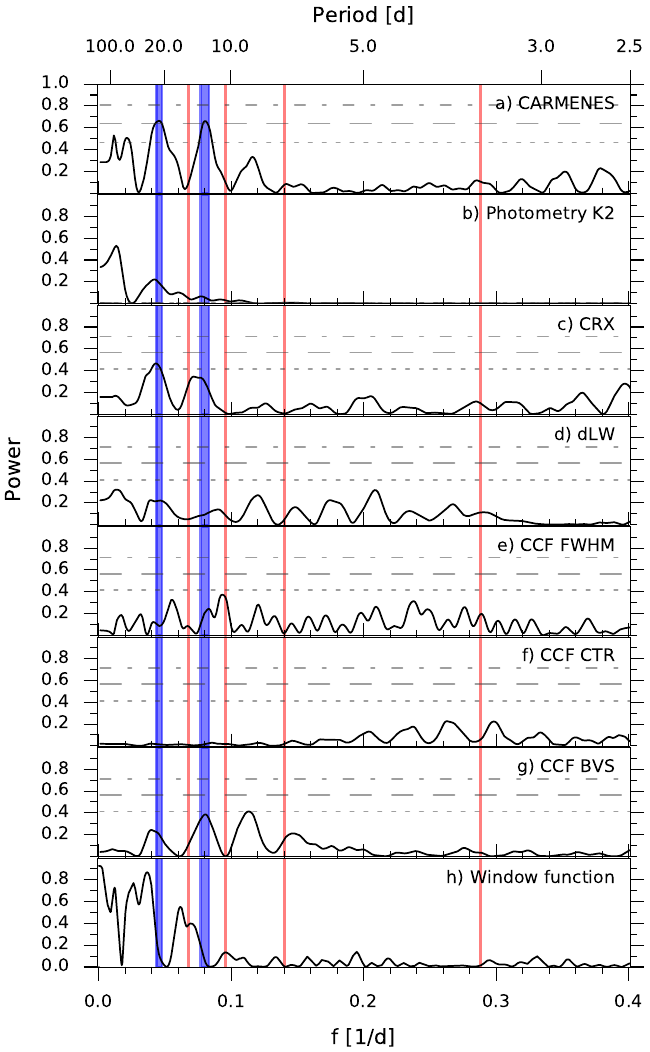}
\caption{From top to bottom: Generalized Lomb-Scargle periodograms (GLS) of the EPIC~246471491 radial velocities from CARMENES data (a), the {\it K2} photometry (b), and the CRX (c), dLW (d), FWHM (e), CTR (f), and BVS (g) indices. The lower panel (h) shows the window function of the data. Vertical red lines indicate the frequencies associated with each of the four transiting planets, and  the blue vertical lines mark the frequencies associated to the activity of the host star. The highest peaks in the CARMENES GLS are at located at $f = 0.045\,\mathrm{d^{-1}}$ ($P \sim 22\,\mathrm{d}$) and $f = 0.081\,\mathrm{d^{-1}}$ ($P = 12.1\,\mathrm{d}$) and are linked to the star's rotation. These periodicities are also significant in the {\it K2} photometry and the CRX index. Horizontal lines show the false alarm probability (FAP) levels of 10\% (short-dashed line), 1\% (long-dashed line) and 0.1\% (dot-dashed line).}
\label{lsp1}
\end{figure}

We performed a frequency analysis of the available radial velocity observations. In Figure~\ref{lsp1} we plot the generalized Lomb-Scargle periodogram (GLS, \citet{GLS}) of the CARMENES radial velocity data. 
For each periodogram, we compute the theoretical false alarm probability (FAP) and mark the 10\%, 1\%, and 0.1\% significance level.  The vertical red lines mark the orbital frequencies of planets b, c, d and e, and the thick blue lines mark the stellar rotational frequencies associated to stellar variability.

It is easily seen that the dominant signals in the CARMENES periodogram are those at $12.1$ days and $\sim 22$ days. These periodicities are also significant in the photometric data from {\it K2} and in the chromatic index (CRX), an indicator developed for CARMENES data to recognize wavelength-dependent variability attributable to stellar activity \citep{SERVAL}. The periodicities are also present, but with $\mathrm{FAP}>10\%$, in the CARMENES differential line width (dLW), CCF bisector velocity span (BVS) and CCF full width half maximum (FWHM) indices. Based solely on the data available to us, it is not clear which one of the two periods is the true rotational period of the star, and which one is indeed an alias or an harmonic of the other. 

There is evidence that $\sini$ should be in fact close to unity for these types of systems \citep{2017AJ....154..270W}. A simple calculation ($P_\mathrm{rot} \sini = 2\pi R \vsini$), using the stellar \vsini\, value and assuming $\sini = 1$, gives an expected stellar rotational period upper limit of $10.2^{+2.6}_{-1.7}$ days. Therefore, we adopt 12.1 days as the true rotational period of the star, $P_\mathrm{rot}$, seen in the CARMENES GLS periodogram. The $\sim 22$ days peak is then an alias originated from the window function peak at $\sim 27.5$ days, caused by our scheduling of optimal observations along the lunar cycle.

\begin{figure}
\centering
\includegraphics[width=8.3cm,angle=0,clip=true]{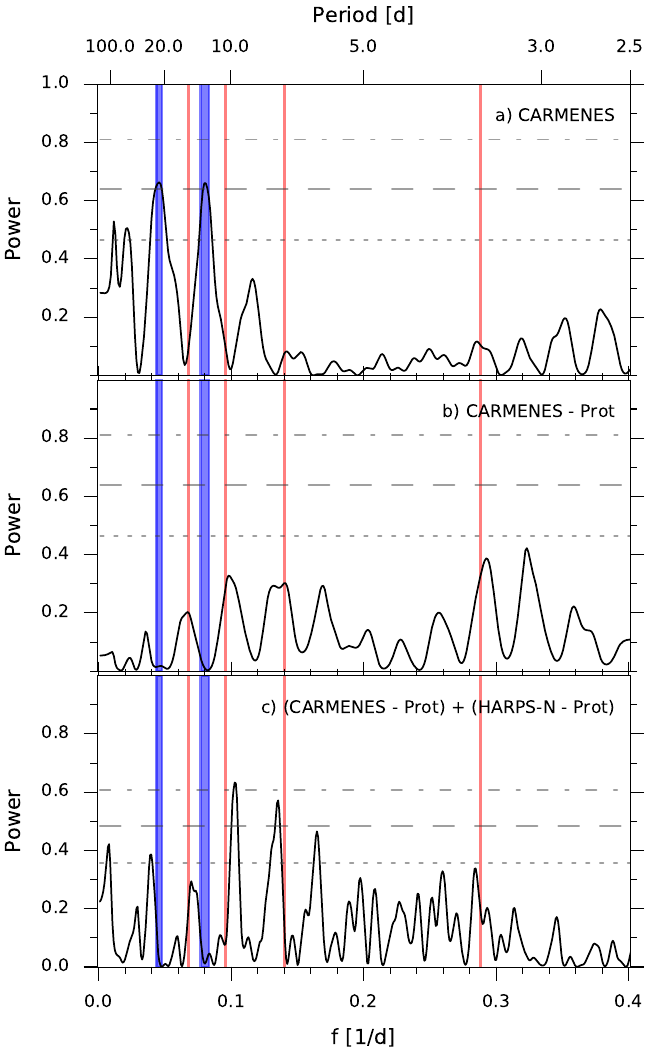}
\caption{Generalized Lomb-Scargle periodograms (GLS) of the EPIC~246471491 radial velocities from CARMENES and HARPS-N. In panel (a) the same data and analysis is shown as panel (a) of Figure~\ref{lsp1}. Panel (b) shows the GLS of the CARMENES data after removing the prominent peak at $f = 0.081\,\mathrm{d^{-1}}$, corresponding to $P_\mathrm{rot} = 12.1$~days. In this case the frequency associated with the alias peak at $\sim 22$ days also loses all the power, and the higher peaks are located at the periodicities of the four planets. In panel (c) the HARPS-N data (also $P_\mathrm{rot}$ corrected), is added to the CARMENES data. The joint GLS shows significant periodicities for 2 of the 4 planets in this system. As in Figure~\ref{lsp1}, horizontal lines show the false alarm probability (FAP) levels of 10\% (short-dashed line), 1\% (long-dashed line) and 0.1\% (dot-dashed line).}
\label{test2}
\end{figure}

In light of these results, it is clear that the dominant signal in the RV data is that of stellar activity, and that this needs to be taken into account in order to retrieve the planetary masses. In Figure~\ref{test2} we show the GLS periodograms of the CARMENES data with the stellar and planetary periodicities marked, the spectral power being dominated by the former. In the middle panel, we filter the data by removing the $P_\mathrm{rot}$ periodicity. We do this by fitting the amplitude and phase of a sinusoidal signal, and computing the GLS periodogram of the residuals of this fit, in the same way as it is done for planetary signals.

This procedure eliminates both the 12.1 days and 22 days signals, thus confirming that the latter is in fact an alias. Now the major peaks in the power spectra correspond to the planetary orbital periods, although they are not above the $\mathrm{FAP}=10\%$ level. In the bottom panel, the HARPS-N data is added accounting for the RV offset between both instruments using the measurements taken in consecutive days with HARPS-N ($\mathrm{JD} \sim 2458046.5$) and CARMENES ($\mathrm{JD} \sim 2458047.5$). Removing $P_\mathrm{rot}$ signal from HARPS-N data does not carry a strong effect on the final result. Regardless, for the sake of consistency, we have removed it in our analysis. A possible explanation may lie in the fact that there is only a handful of measurements (9), or that the HARPS-N and CARMENES spectrographs cover different spectral ranges and thus their sensitivity to stellar activity is different. The GLS periodogram of the combined data shows significance peaks ($\mathrm{FAP} \approx 0.1\%$) at the orbital periods of planets c and d, and above $\mathrm{FAP}=10\%$ for planets b and e.

\section{Joint analysis and mass determinations}

In order to retrieve the masses of the planets in the EPIC~246471491 system, we performed a joint analysis of the photometric {\it K2} data and the radial velocity measurements from CARMENES and HARPS-N. We make use of the Pyaneti\footnote{\href{https://github.com/oscaribv/pyaneti}{https://github.com/oscaribv/pyaneti}} code \citep{2016AJ....152..193B}, which uses Markov chain Monte Carlo (MCMC) techniques to infer posterior distributions for the fitted parameters. The radial velocity data are fitted with Keplerian orbits, and we use the limb-darkened quadratic transit model by \citet{2002ApJ...580L.171M} to fit the transit light curves. These methods have already been successfully applied to several planets, see for example \citet{2017AJ....154..266N} or \citet{jorge} for details.

Although no coherent rotational modulation is found in the {\it K2} data alone, as in \citet{jorge}, the light curve of EPIC~246471491 suggests that the evolution time scale of active regions is longer than the {\it K2} observations (80 days). Since our combined observations cover 112 days, we decided to model the stellar activity signal with a sinusoid \citep{2013Natur.503..377P, 2017arXiv171102097B}. Thus, on top of the planetary signals, we include in the fit a fifth radial velocity signal corresponding to the stellar variability at $P_\mathrm{rot}$, which we identified as the dominant RV signal in the previous section. \citet{2015ApJ...808..126V} reported that the eccentricity of small planets in {\it Kepler} multi-planet systems is low. Given that EPIC~246471491 is a compact short-period multi-planetary system, we also assumed tidal circularization of the orbits and fixed the eccentricity to zero for all four planets.

\begin{figure}
\centering
\includegraphics[width=8.3cm,angle=0,clip=true]{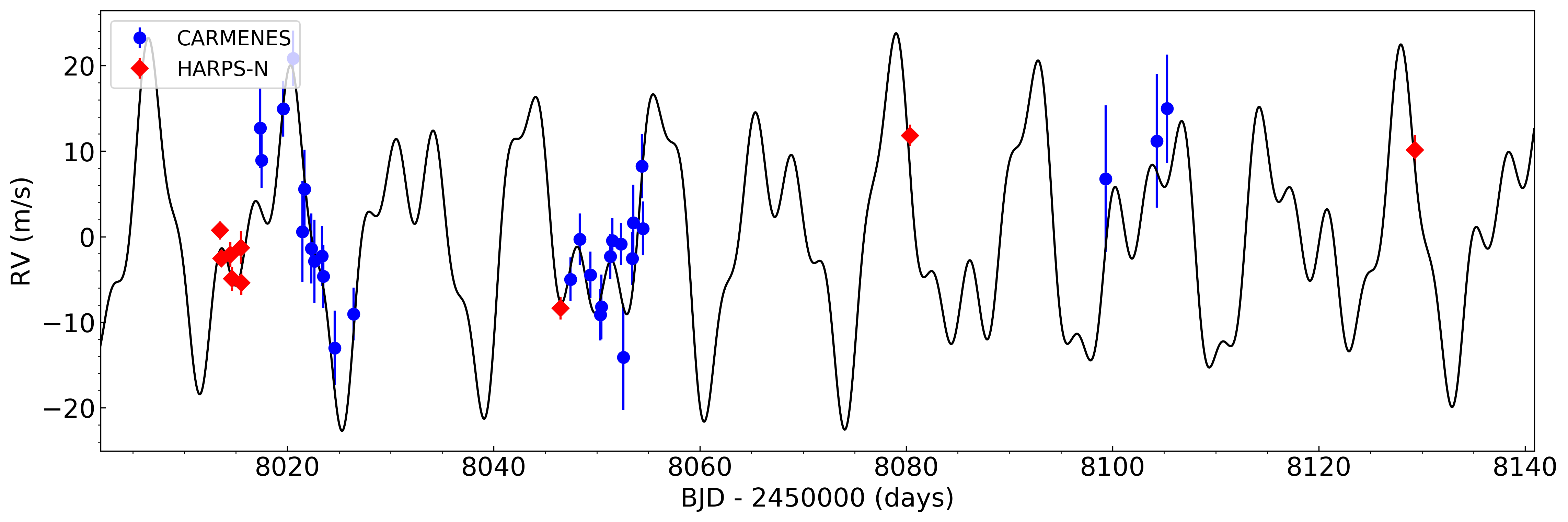}
\caption{Time series of the RV measurements of EPIC~246471491 derived from CARMENES (blue dots) and HARPS-N (red diamonds). The black line corresponds to the best-fit model to the data, which includes the RV signal of each of the four planets and the stellar activity. }
\label{timeseries}
\end{figure}

Figure~\ref{timeseries} shows the combined CARMENES and HARPS-N radial velocity measurements plotted against time, with a superimposed best-fit model containing the radial velocity variations due to the four planets and a stellar activity signal. In our analysis, we did not discard RV observations that were obtained during transits, but the expected Rossiter-McLaughlin amplitude is negligible.

\begin{figure*}
\centering
\includegraphics[width=8.3cm,angle=0,clip=true]{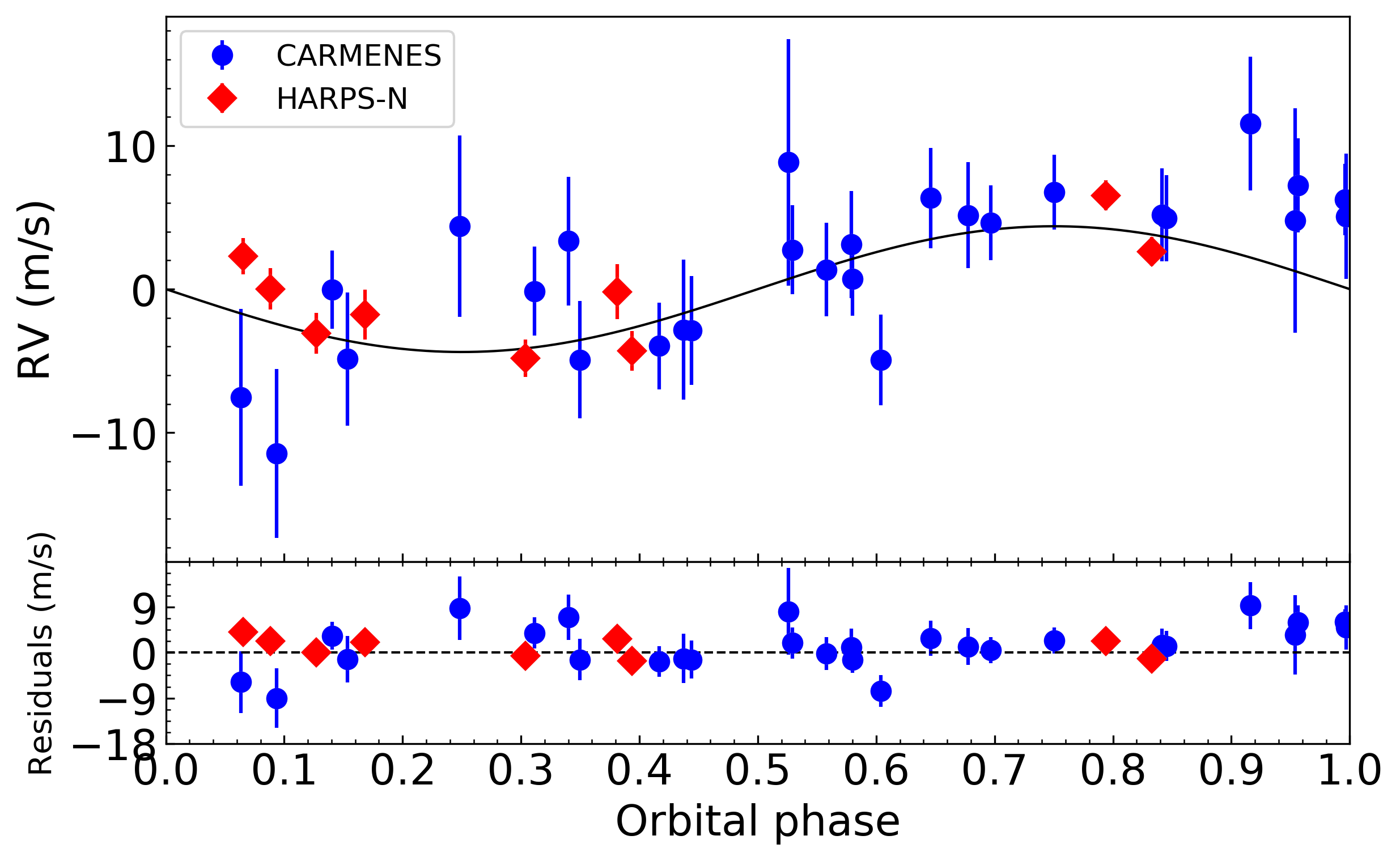}
\includegraphics[width=8.3cm,angle=0,clip=true]{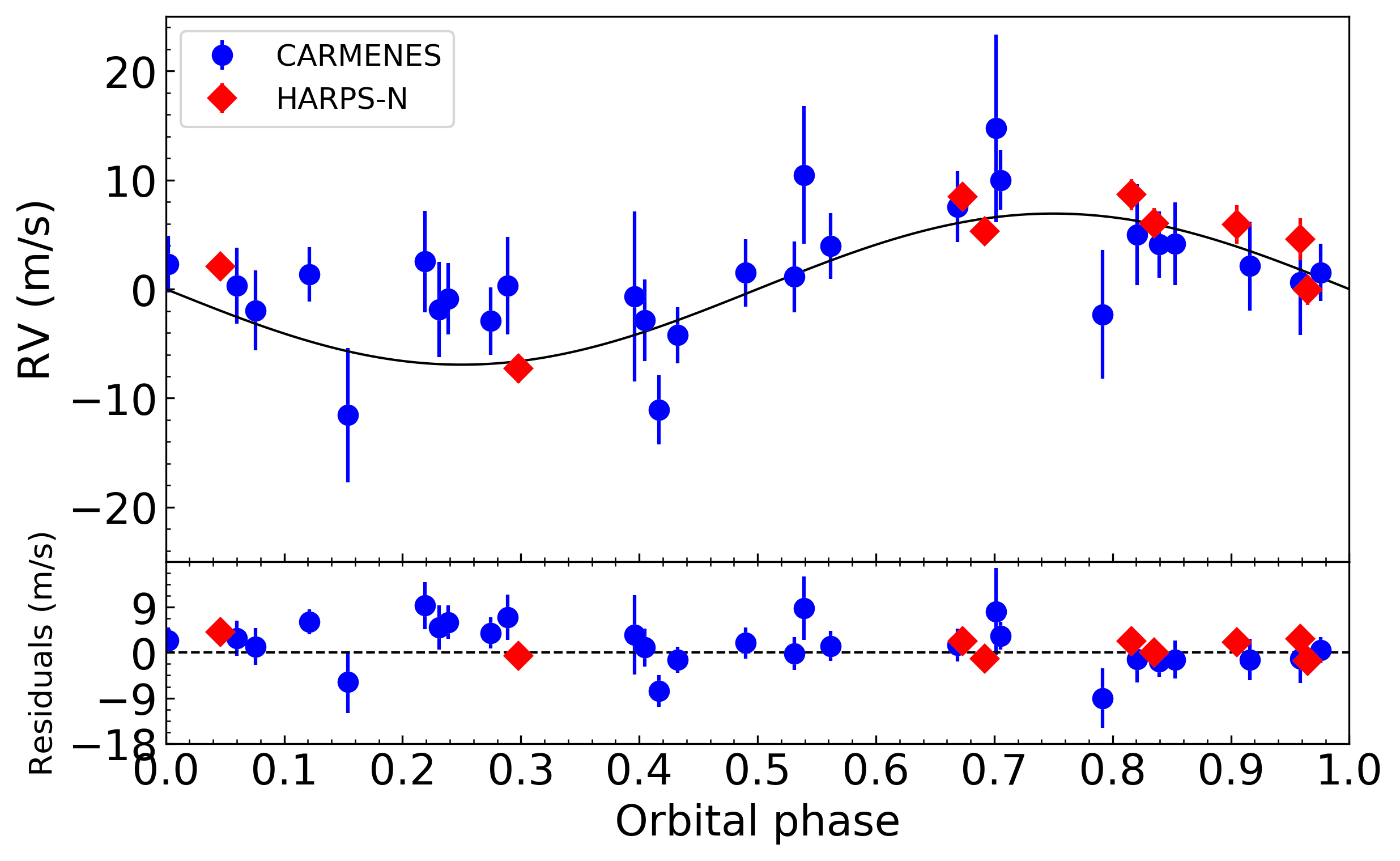}
\includegraphics[width=8.3cm,angle=0,clip=true]{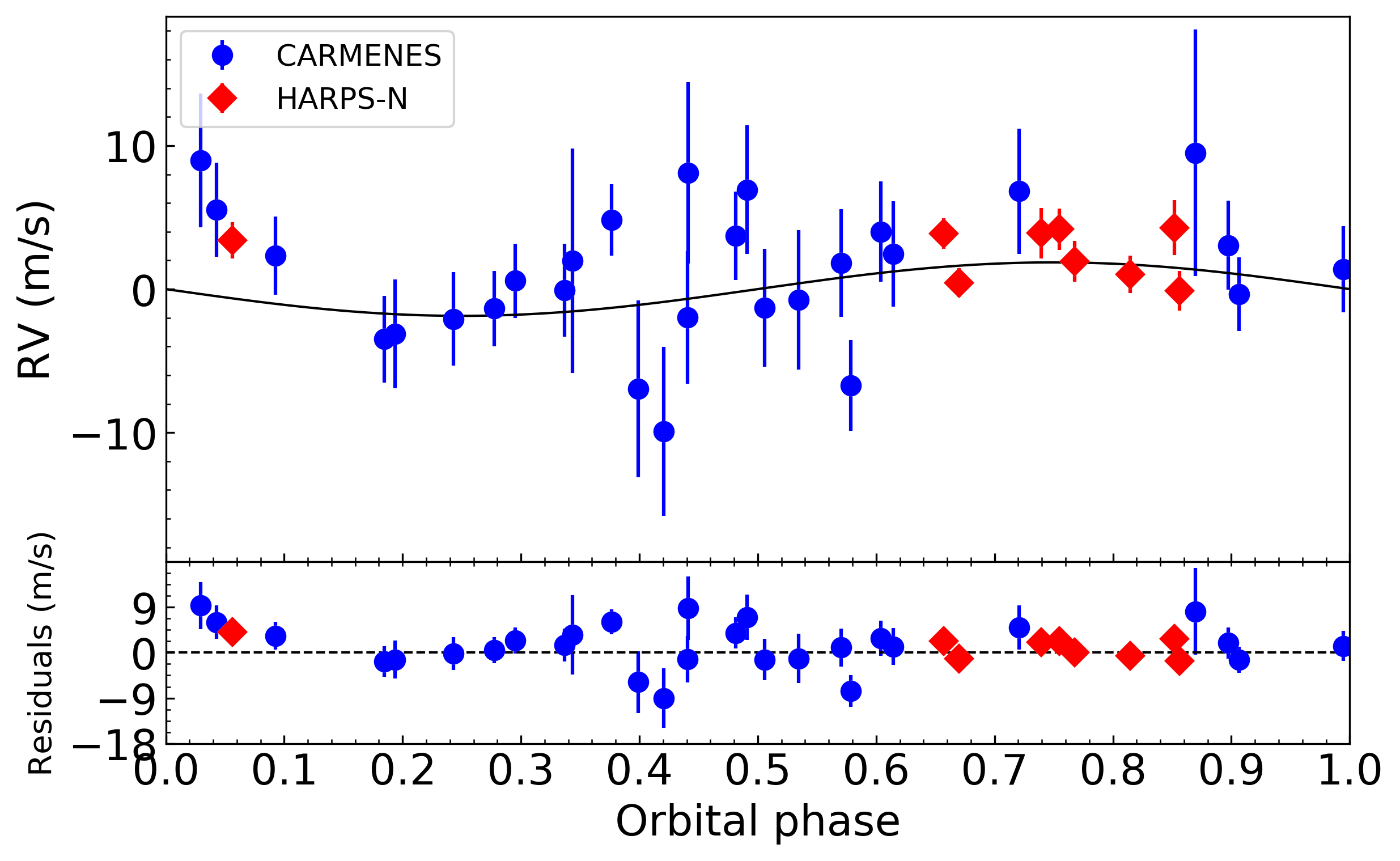}
\includegraphics[width=8.3cm,angle=0,clip=true]{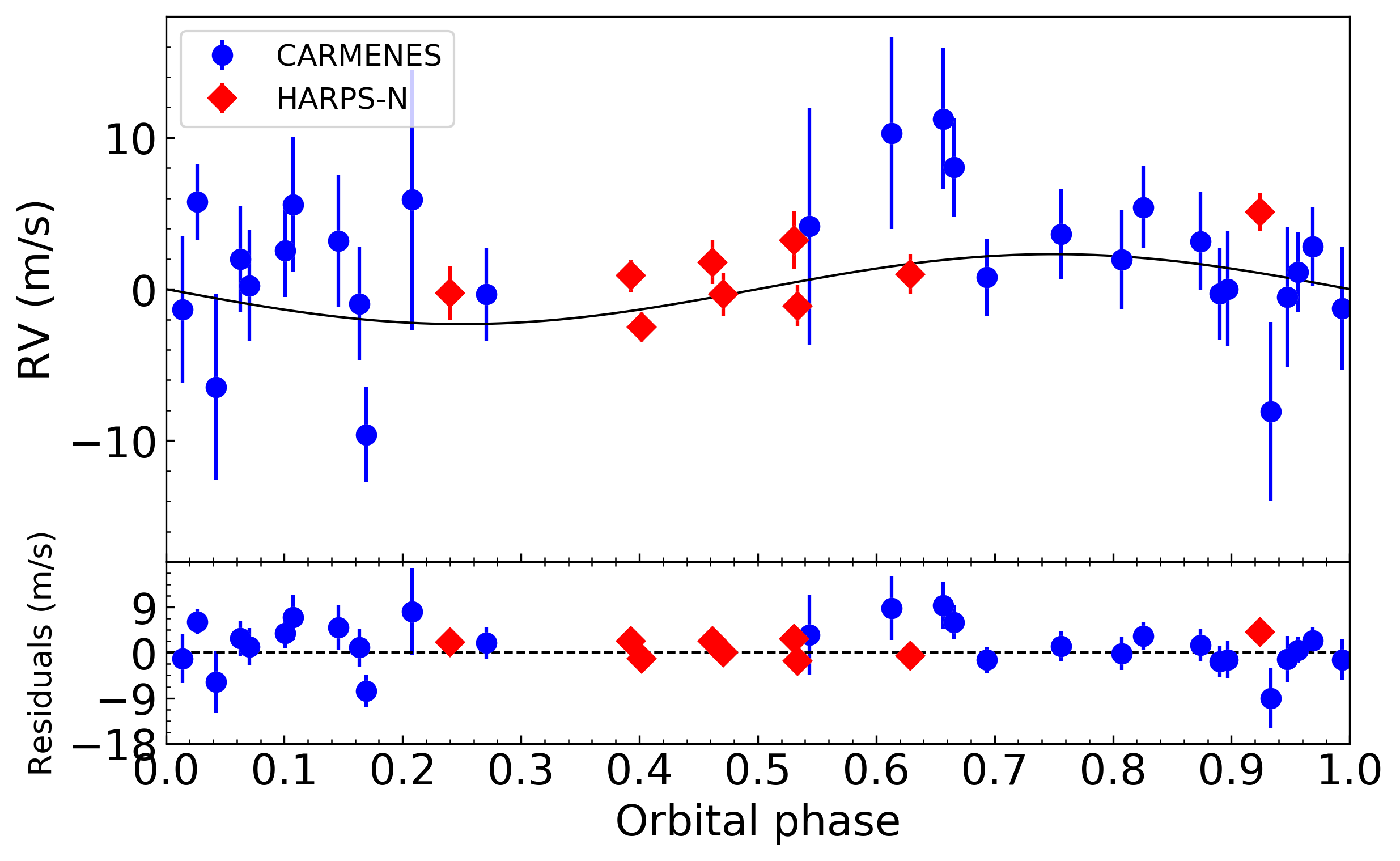}
\includegraphics[width=8.3cm,angle=0,clip=true]{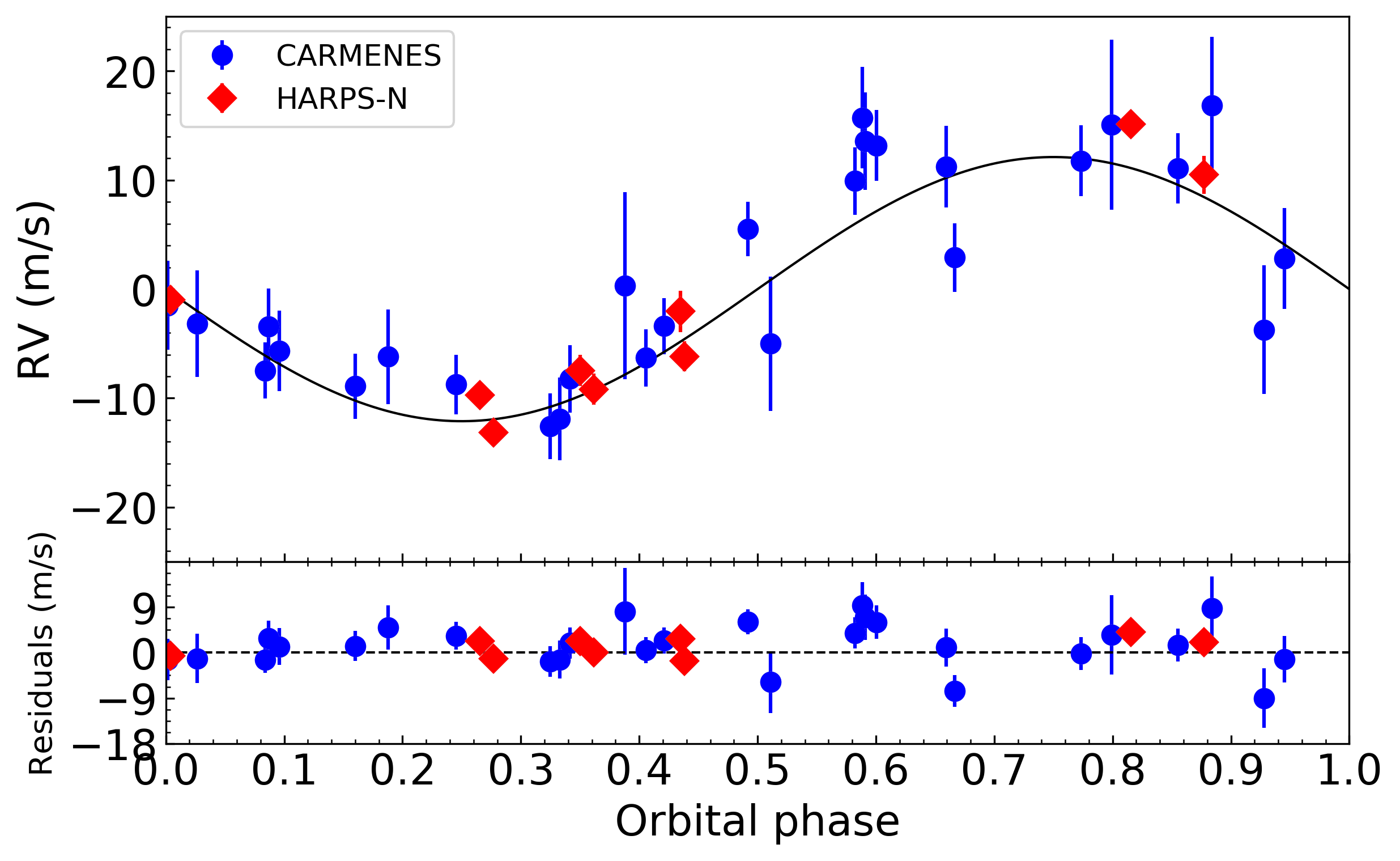}
\caption{{\it Top left}: Phase-folded radial velocity measurements of EPIC~246471491 over the period of planet b, after eliminating the signal from planets c, d, e, and the stellar activity signal. {\it Top right}: Same for planet c. {\it Middle left}: Same for planet d. {\it Middle right}: Same for planet e. {\it Bottom}: Same for the stellar activity signal. The derived semi-amplitudes for these planets are  \kb[]\,$\mathrm{m\,s^{-1}}$,  \kc[]\,$\mathrm{m\,s^{-1}}$,  \kd[]\,$\mathrm{m\,s^{-1}}$ and \ke[]\,$\mathrm{m\,s^{-1}}$, while the stellar activity signal has an amplitude of \kf[]\,$\mathrm{m\,s^{-1}}$. This translates into a mass determination of \mpb[]\,$M_\oplus$ and \mpc[]\,$M_\oplus$ for planets b and c, respectively, and an upper limit mass determination of 6.5\,$M_\oplus$ and 10.7\,$M_\oplus$ (at 99\% confidence level) for planets d and e, respectively.}
\label{phase2}
\end{figure*}

The individual phased RV signals for each of the four planets, once the stellar variability signal and the other planet's signals have been removed, are shown in Figure~\ref{phase2}. Also shown is the phased stellar activity signal once the four planet's signals have been removed. The residuals around the best fit model are shown below each panel. The radial velocity signal of the stellar activity is readily detectable and has the largest RV semi-amplitude. We can also identify at larger than $3 \sigma$ significance level the semi-amplitudes of planets b and c, while we can only place upper limits to the masses of planets d and e. In Table~\ref{planetparams} the planet properties of the EPIC~246471491 system are summarized.

As a further test, we used the code \texttt{SOAP2} \citep{Dumusque2014} to estimate the expected induced RV signal coming from stellar activity. We assume that spots generate a flux decrement of 1.5\% from the largest depth in the light curve (Figure \ref{lightcurve}). We used the stellar parameters from Table~\ref{table_star} and a stellar rotation period of 12.3 days. We assume that the star has two spots separated by 180 deg located at the stellar equator. \texttt{SOAP2}'s output gives an expected induced RV signal of $\sim 13\,{\rm m\,s^{-1}}$. This result is consistent with the fitted amplitude in our model.

\section{Discussion and Conclusions}
\label{sec:discuss}

\begin{figure}
\centering
\includegraphics[width=8.3cm,angle=0,clip=true]{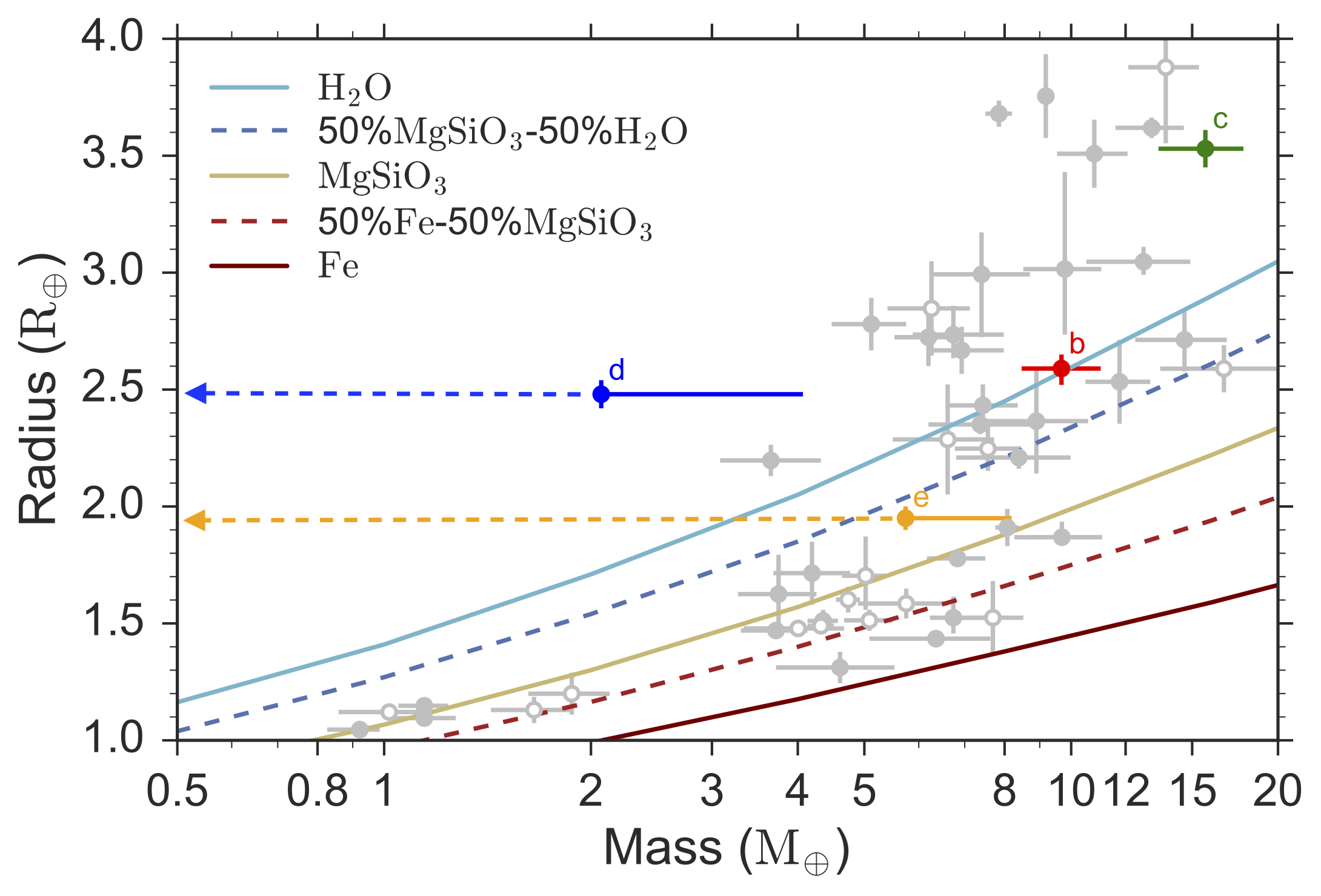}
\caption{Mass-radius diagram for all known planets with masses between 0.5--20\,$M_\oplus$ and radius 1--4\,$R_\oplus$, comprising from Earth-like to super-Earth to sub-Neptune regimes. Data are taken from the TEPCat database \citep{2011MNRAS.417.2166S}. Planets belonging to multiple systems are marked in open gray dots while single planets are marked with solid gray dots. The four planets of EPIC~246471491's system are marked in different colors. Theoretical models for the planet's internal composition are taken from \citet{2016ApJ...819..127Z}. }
\label{comparativa}
\end{figure}

We determined masses, radii, and densities for two of the four planets known to transit EPIC~246471491. We find that EPIC~246471491~b has a mass of $M_\mathrm{b}$\,=\,\mpb\, and a radius of $R_\mathrm{b}$\,=\,\rpb, yielding a mean density of $\rho_\mathrm{b}$\,=\,\denpb, while EPIC~246471491~c has a mass of $M_\mathrm{c}$\,=\,\mpc, radius of $R_\mathrm{c}$\,=\,\rpc, and a mean density of $\rho_\mathrm{c}$\,=\,\denpc. For EPIC~246471491~d and EPIC~246471491~e we are able to calculate upper limits for the masses at $6.5\,M_\oplus$ and $10.7\,M_\oplus$, respectively. 

\citet{Fulton17} and \citet{van2017} reported a bi-modal distribution in the radii of small planets at the boundary between super-Earths and sub-Neptunes. A clear distinction between two different families of planets is reported: on the one hand super-Earths have a radius distribution that peaks at $R_\mathrm{p}$\,$\sim$\,1.5\,$R_\oplus$, and on the other sub-Neptune planets have a radius distribution that peaks at $R_\mathrm{p}$\,$\sim$\,2.5\,$R_\oplus$. These two populations are separated by a gap in the radius distribution.

Figure~\ref{comparativa} illustrates the mass-radius diagram of all known planets with precise mass determination, extending the full parameter space encompassing Earth-like, super-Earth and Neptune planets (1--4\,$R_\oplus$, 0.5--20\,$M_\oplus$). The four planets of the EPIC~246471491 system are also plotted. Two of the planets, b and d, fall in the sub-Neptune category, with radius very close to one of the peaks of the bi-modal distribution at 2.5 $R_\oplus$, while planet e belongs to the scarce population of planets located within the radius gap. Planet c is a larger object with only a slightly smaller radius and larger density than Neptune (3.9 vs. 3.5\,$R_\oplus$ and 1.64 vs. 1.95\,$\mathrm{g\,cm^{-3}}$; for Neptune and EPIC~246471491~c, respectively).

Using the values in Table~\ref{planetparams}, the estimated transmission signals corresponding to H/He atmospheres (which would be the optimistic case for super-Earth size planets) of the four planets would be of 20, 32, 21 and 8\,ppm for planets b, c, d and e, respectively. For planets d and e the upper mass limit has been used for the calculations, so presumably the true signals would be larger. Still, with such relatively small atmospheric signatures, the planets are not optimal for transmission spectroscopy studies using current instrumentation due to the faintness of the parent star. 

However, as in the case of the triple transiting system {\it K2}-135 \citep{2017AJ....154..266N, jorge}, the four planets around EPIC~246471491 could provide a great test case to study comparative atmospheric escape and evolution within the same planetary system. From Figure~\ref{comparativa} it is readily seen that the two planets with well determined mass, have very different densities. Planet b has a bulk density close to pure water, while planet c is a much more inflated lower density planet. Assuming that all planets in the system were formed with similar composition, the different bulk densities could be explained by the factor 5 larger insolation flux received by planet b, compared to c, driving atmospheric escape and mass loss. While the masses of the other two planets are only upper limits, planet d (the third in distance from the star) clearly falls in the low density regime, which would be consistent with this hypothesis. For planet e, a larger range in densities is possible, from pure MgSiO$_3$ to extremely low densities. Thus, comparative studies focused on exosphere and atmospheric escape processes, through the detection of H$\alpha$, Ly$\alpha$, or He lines can be conducted for EPIC~246471491 with the next-generation of Extremely Large Telescopes (ELTs).

\begin{acknowledgements}
     CARMENES is an instrument for the Centro Astron\'omico Hispano-Alem\'an de Calar Alto (CAHA, Almer\'{\i}a, Spain). CARMENES is funded by the German Max-Planck-Gesellschaft (MPG), the Spanish Consejo Superior de Investigaciones Cient\'{\i}ficas (CSIC), the European Union through FEDER/ERF FICTS-2011-02 funds, and the members of the CARMENES Consortium (Max-Planck-Institut f\"ur Astronomie, Instituto de Astrof\'{\i}sica de Andaluc\'{\i}a, Landessternwarte K\"onigstuhl, Institut de Ci\`encies de l'Espai, Insitut f\"ur Astrophysik G\"ottingen, Universidad Complutense de Madrid, Th\"uringer Landessternwarte Tautenburg, Instituto de Astrof\'{\i}sica de Canarias, Hamburger Sternwarte, Centro de Astrobiolog\'{\i}a and Centro Astron\'omico Hispano-Alem\'an), with additional contributions by the Spanish Ministry of Economy, the German Science Foundation through the Major Research Instrumentation Programme and DFG Research Unit FOR2544 ``Blue Planets around Red Stars'', the Klaus Tschira Stiftung, the states of Baden-W\"urttemberg and Niedersachsen, and by the Junta de Andaluc\'{\i}a.
   This article is based on observations made in the Observatorios de Canarias del IAC with the TNG telescope operated on the island of La Palma by the Galileo Galilei Fundation, in the Observatorio del Roque de los Muchachos (ORM). HARPS-N data were taken under observing programs CAT17A-91, A36TAC-12 and OPT17B-59.
    This work is partly financed by the Spanish MINECO through grants ESP2016-80435-C2-1-R, ESP2016-80435-C2-2-R and  AYA2016-79425-C3-3-P. This work was also supported by JSPS KAKENHI Grants Numbers JP16K17660 and JP18H01265.
  
\end{acknowledgements}

\bibliographystyle{aa} 
\bibliography{ref_db} 

\clearpage

\begin{table}[!t]
\begin{center}
\caption{Radial velocity measurements derived from HARPS-N and CARMENES observations used in this paper.}
\label{rvs_carmenes}
\begin{tabular}{lccc}
\hline
  JD           & RV [m\,s$^{-1}$]          & Error [m\,s$^{-1}$]      & Instrument \\
\hline
  2458013.4565 & 2.72 & 1.06 & HARPS-N\\
  2458013.59056 & -0.57 & 1.01 & HARPS-N\\
  2458014.47738 & -0.13 & 1.44 & HARPS-N\\
  2458014.61324 & -2.95 & 1.42 & HARPS-N\\
  2458015.49584 & 0.68 & 1.92 & HARPS-N\\
  2458015.53839 & -3.43 & 1.38 & HARPS-N\\
  2458017.35179 & 12.56 & 4.67 & CARMENES\\
  2458017.49193 & 8.79 & 3.27 & CARMENES\\
  2458019.58022 & 14.81 & 3.26 & CARMENES\\
  2458020.56477 & 20.69 & 3.24 & CARMENES\\
  2458021.44042 & 0.44 & 5.90 & CARMENES\\
  2458021.64733 & 5.40 & 4.63 & CARMENES\\
  2458022.3294 & -1.55 & 4.09 & CARMENES\\
  2458022.63247 & -3.02 & 4.87 & CARMENES\\
  2458023.35717 & -2.41 & 3.49 & CARMENES\\
  2458023.46814 & -4.80 & 3.69 & CARMENES\\
  2458024.57625 & -13.15 & 4.36 & CARMENES\\
  2458026.42467 & -9.19 & 3.09 & CARMENES\\
  2458046.47123 & -6.38 & 1.32 & HARPS-N\\
  2458047.43121 & -5.17 & 2.57 & CARMENES\\
  2458048.35116 & -0.46 & 3.01 & CARMENES\\
  2458049.37702 & -4.61 & 2.72 & CARMENES\\
  2458050.33568 & -9.29 & 3.02 & CARMENES\\
  2458050.43032 & -8.37 & 3.80 & CARMENES\\
  2458051.30778 & -2.47 & 2.62 & CARMENES\\
  2458051.49481 & -0.60 & 2.60 & CARMENES\\
  2458052.34674 & -1.01 & 2.48 & CARMENES\\
  2458052.58013 & -14.27 & 6.17 & CARMENES\\
  2458053.4422 & -2.72 & 3.09 & CARMENES\\
  2458053.54225 & 1.46 & 4.47 & CARMENES\\
  2458054.37099 & 8.08 & 3.74 & CARMENES\\
  2458054.45669 & 0.80 & 3.16 & CARMENES\\
  2458080.36219 & 13.81 & 1.27 & HARPS-N\\
  2458099.31923 & 6.61 & 8.59 & CARMENES\\
  2458104.27631 & 11.03 & 7.81 & CARMENES\\
  2458105.29813 & 14.83 & 6.32 & CARMENES\\
  2458129.3238 & 12.10 & 1.75 & HARPS-N\\
\hline
\end{tabular}
\end{center}
\end{table}

\begin{table*} 
\tiny
  {\renewcommand{\arraystretch}{1.2}
  \caption{Summary of the system parameters of EPIC~246471491 determined in section \ref{sec:photo} using only the fit to the photomtric data from {\it K2} mission, and in section \ref{sec:results} with the Pyaneti code fitting simultaneously the photometric and radial velocity data. }  
  \begin{tabular}{lccccc}
  \hline
  \hline
  \noalign{\smallskip}
  Parameter & EPIC~246471491~b & EPIC~246471491~c & EPIC~246471491~d & EPIC~246471491~e &  Stellar signal \\
  \noalign{\smallskip}
  \hline
  \noalign{\smallskip}
  \noalign{\smallskip}
  \multicolumn{3}{l}{\emph{\bf Model fits to {\it K2} data only}} \\
  \noalign{\smallskip}
Orbit inclination $i_\mathrm{p}$ ($^{\circ}$)                 & $87.0^{+2.0}_{-2.0}$      & $88.0^{+1.0}_{-1.0}$                   & $89.2^{+0.6}_{-0.9}$            & $89.4^{+0.4}_{-0.6}$     &      \\
Semi-major axis $a$ $(R_{*})$       & $11.0^{+2.0}_{-3.0}$    & $17.0^{+4.0}_{-4.0}$                    & $30.0^{+3.0}_{-6.0}$            & $45.0^{+5.0}_{-10.0}$     &      \\
 Transit epoch $T_0$ (JD$-$2\,454\,833)      & $2910.3753^{+0.0006}_{-0.0007}$  & $2911.5384^{+0.0004}_{-0.0007}$ &    $2912.201^{+0.001}_{-0.001}$     & $2908.897^{+0.003}_{-0.002}$       &    \\
Planet radius $R_\mathrm{p}$ ($R_{\rm \oplus}$)        & $2.62^{+0.05}_{-0.04}$ & $3.7^{+0.3}_{-0.3}$                      & $2.57^{+0.09}_{-0.06}$          & $2.01^{+0.20}_{-0.09}$       &    \\
Orbital period $P_{\mathrm{orb}}$ (days)                     & $3.47175^{0.00004}_{-0.00005}$   & $7.13804^{+0.00007}_{-0.00010}$ & $10.4560^{ 0.0004}_{-0.0003}$  & $14.7634^{0.0007}_{-0.0006}$        &   \\
Impact parameter $b$                   & $0.5^{+0.8}_{-1.0}$   & $0.5^{+0.7}_{-0.7}$                      & $0.4^{+0.3}_{-0.4}$             & $0.5^{+0.2}_{-0.3}$        &   \\
Transit depth                                     & $0.00097^{+0.00004}_{-0.00003}$ & $0.00210^{+0.0003}_{-0.0002}$   & $0.00100^{+0.00007}_{-0.00004}$ & $0.00068^{+0.00009}_{-0.00006}$      &     \\
Transit duration $\tau_{14}$ (hours)                 & $2.57\,\pm\,0.02$  & $3.08\,\pm\,0.03$                          & $2.96\,\pm\,0.05$              & $2.76\,\pm\,0.07$       &    \\
Linear limb-darkening coefficient $u_1$                   & $0.2^{+0.2}_{-0.1}$  & $0.5^{+0.3}_{-0.2}$                       & $0.3^{+0.4}_{-0.3}$             & $0.9^{+0.5}_{-0.5}$      &     \\
Quadratic limb-darkening coefficient $u_2$                  & $0.06^{+0.07}_{-0.05}$    & $0.38^{+0.10}_{-0.09}$                & $0.18^{+0.10}_{-0.09}$          & $0.0^{+0.3}_{-0.2}$      &     \\
Eccentricity$^{(a)}$ $e$          &                     0           \\
Longitude of periastron$^{(a)}$ $\omega_{\star}$ ($^{\circ}$)      & 90           \\
    \noalign{\smallskip}
  \hline
  \noalign{\smallskip}
  \noalign{\smallskip}
  \multicolumn{3}{l}{\emph{\bf Model Parameters: Pyaneti}} \\
  \noalign{\smallskip}
    Orbital period $P_{\mathrm{orb}}$ (days)  & \Pb[] & \Pc[] & \Pd[] & \Pe[] & \Pf[]\\ 
    Transit epoch $T_0$ (JD$-$2\,450\,000)  & \Tzerob[] & \Tzeroc[] & \Tzerod[] & \Tzeroe[] & \Tzerof[]\\  
    Scaled planet radius $R_\mathrm{p}/R_{\star}$ & \rrb[] & \rrc[] & \rrd[] & \rre[]& $\cdots$  \\
    Impact parameter $b$ & $0.57^{+0.08}_{-0.15}$  & $0.05^{+0.04}_{-0.04}$ & $0.20^{+0.04}_{-0.05}$            & $0.17^{+0.04}_{-0.07}$     &    \\
$\sqrt{e} \sin \omega_\star^{(a)}$  & 0 & 0 & 0& 0 \\
    $\sqrt{e} \cos \omega_\star^{(a)}$  & 0 & 0 & 0 & 0\\
Doppler semi-amplitude $K$ (m s$^{-1}$) & \kb[]  &  \kc[]  &  \kd[] &  \ke[]  &  \kf[] \\
    Systemic velocity $\gamma_{\rm CARMENES}$ (km\,s$^{-1}$)  & \CARMENES[] \\
    Systemic velocity $\gamma_{\rm HARPS-N}$ (km\,s$^{-1}$)  & \HARPSN[] \\
    Limb-darkening coefficient $q_1^{(b)}$  &  \qone  \\
    Limb-darkening coefficient $q_2^{(b)}$  & \qtwo \\  
    \noalign{\smallskip}
        \hline
    \noalign{\smallskip}
    \multicolumn{3}{l}{\emph{\bf Derived Parameters: Pyaneti}} \\
    \noalign{\smallskip}
    Planet mass $M_\mathrm{p}$ ($M_{\rm \oplus}$) &  \mpb[] &  \mpc[] &   $ <6.5$       &   $< 10.7$      \\        
    Planet radius $R_\mathrm{p}$ ($R_{\rm \oplus}$)  & \rpb[]  & \rpc[] & \rpd[] & \rpe[]   \\
    Planet density $\rho_{\rm p}$ (g\,cm$^{-3}$)  & \denpb[] & \denpc[]   \\   
    Surface gravity $g_{\rm p}$ (cm\,s$^{-2}$)  & \grapparsb[] & \grapparsc[] \\   
    Surface gravity$^{(c)}$ $g_{\rm p}$ (cm\,s$^{-2}$)  & \grapb[] & \grapc[] \\   
	Scaled semi-major axis $a/R_\star$  & \arb[] & \arc[] & \ard[]& \are[]  \\
	Semi-major axis $a$ (AU)  & \ab[] & \ac[]& \ad[]& \aeee[]  \\
    Orbit inclination $i_\mathrm{p}$ ($^{\circ}$) & \ib[] & \ic[] & \id[] & \ie[] \\
    Transit duration $\tau_{14}$ (hours) & \ttotb[] & \ttotc[] & \ttotd[] & \ttote[]\\
    Equilibrium temperature$^{(\mathrm{d})}$  $T_\mathrm{eq}$ (K)  & \Teqb[] & \Teqc[] & \Teqd[] & \Teqe[]\\
        Insolation  $F$ ($F_{\oplus}$)  & \insolationb[] & \insolationc[] & \insolationd[] & \insolatione[] \\
    Stellar density (from light curve) & \denstrb[] \\
    Linear limb-darkening coefficient $u_1$ & \uone \\
    Quadratic limb-darkening coefficient $u_2$ &  \utwo\\
    \noalign{\smallskip}
\hline
 \end{tabular}
 \begin{tablenotes}\footnotesize
\item \emph{Note} -- $^{(\mathrm{a})}$ Fixed. $^{(\mathrm{b})}$ $q_1$ and $q_2$ as defined by \citet{kipping13}. $^{(c)}$ Calculated from the scaled parameters as described by \citet{winn10}. $^{(d)}$~Assuming albedo = 0. 
\end{tablenotes}
\label{planetparams}}
\end{table*}







\end{document}